\documentclass[12pt,a4paper]{article} 

\usepackage{graphicx}

\usepackage{epsfig}
\usepackage{cite}

\usepackage{color} % for comments
\usepackage{amssymb}
\usepackage{amsmath}
\usepackage{authblk}
\usepackage{doi}
\usepackage{dsfont}
\usepackage{setspace}
\usepackage{subfloat}
\usepackage{subfig}
\usepackage{float}
\usepackage{mathtools}
\usepackage{verbatim}

\usepackage{slashed}% for missing E_T or missing p_T
\usepackage[normalem]{ulem} % for strike through with \sout command, remove it for the final version

\usepackage[dvipsnames]{xcolor}

\newcommand{\mr}{\mathrm}
\newcommand{\mc}{\mathcal}

\newcommand{\wpwpjj}{W^+W^+jj}
\newcommand{\wpzjj}{W^+Zjj}
\newcommand{\wpmzjj}{W^\pm Zjj}
\newcommand{\wzjj}{WZjj}

\newcommand{\zzjj}{ZZjj}

\newcommand{\muf}{\mu_\mr{F}}
\newcommand{\mur}{\mu_\mr{R}}
\newcommand{\xif}{\xi_\mr{F}}
\newcommand{\xir}{\xi_\mr{R}}

\newcommand{\pmmvjj}{\nu_e e^+\mu^-\mu^+jj}

\newcommand{\emmvjj}{\bar\nu_e e^-\mu^-\mu^+jj}

\newcommand{\vevv}{\bar\nu_e e^-\nu_\mu\bar\nu_\mu}
\newcommand{\qqmm}{q\bar q' \mu^-\mu^+}
\newcommand{\qqQQ}{q\bar q' Q \bar Q}
\newcommand{\emQQ}{\bar\nu_e e^-Q\bar Q}
\newcommand{\pmQQ}{\nu_e e^+Q\bar Q}

\newcommand{\smlin}{{\tt SM+lin}}
\newcommand{\smquad}{{\tt SM+quad}}

\newcommand{\VBFNLO}{{\tt{VBFNLO}}}

\newcommand{\MGAMCNLO}{{\tt{MadGraph5\_aMC@NLO}}}

\newcommand{\POWHEG}{{\tt{POWHEG}}}
\newcommand{\POWHEGBOX}{{\tt{POWHEG~BOX}}}
\newcommand{\POWHEGBOXVV}{{\tt{POWHEG~BOX~V2}}}

\newcommand{\PYTHIA}{{\tt{PYTHIA}}}

\newcommand{\PYTHIAE}{{\tt{PYTHIA8}}}

\newcommand{\HERWIGS}{{\tt{HERWIG7}}}

\newcommand{\beq}{\begin{equation}}
\newcommand{\eeq}{\end{equation}}

\newcommand{\bea}{\begin{eqnarray}}
\newcommand{\eea}{\end{eqnarray}}

\begin{document}
\begin{flushright} CERN-TH-2024-036 \end{flushright}
\begin{center}
\section*{QCD effects in electroweak $\wzjj$ production at current and future hadron colliders} 
%\commentbj{Alternative: ``A tool for the simulation of QCD effects in electroweak $\wzjj$ production at current and future hadron colliders''}

\vspace{2.5cm}
{\bf{B.~J\"ager$^1$, A.~Karlberg$^2$, S.~Reinhardt$^1$}}

\end{center}

\vskip 2.5cm

\begin{center}
{\it 
$^1$ Institute for Theoretical Physics, University of T\"ubingen,
Auf der Morgenstelle 14, 72076 T\"ubingen, Germany \\ \noindent
\\[2ex]
$^2$ CERN, Theoretical Physics Department, CH-1211 Geneva 23, Switzerland
\\
}
\noindent
\end{center}
\vfill

\begin{abstract}
{We present an update of an existing implementation of $WZjj$
  production via vector-boson scattering in the framework of the
  \POWHEGBOX{} program. In particular, previously unavailable
  semi-leptonic and fully hadronic decay modes of the intermediate
  vector bosons are provided, and operators of dimension six in an
  effective-field theory approach to account for physics beyond the
  Standard Model in the electroweak sector are included.
For selected applications phenomenological results are provided to
illustrate the capabilities of the new program. The impact of the
considered dimension-six operators on experimentally accessible
distributions is found to be small for current LHC energies, but
enhanced in the kinematic reach of a potential future hadron collider
with an energy of 100~TeV. 
The relevance of fully accounting for spin correlations and off-shell
effects in the decay system is explored by a comparison with results
obtained with the \texttt{MadSpin} tool that are based on an
approximate treatment of the leptonic final state resulting from
vector boson scattering processes.
For selected semi-leptonic and hadronic decay modes we demonstrate the
sensitivity of realistic signal selection procedures on QCD
corrections and parton-shower effects.  }

\end{abstract}

\newpage

\section{Introduction}
Vector boson scattering (VBS) processes are a particularly appealing
class of reactions for exploring the electroweak (EW) sector of the Standard Model
(SM) and possible extensions thereof. 
In the context of the SM, cross sections for the scattering of the longitudinal modes of an EW gauge boson are unitarised by Higgs-boson exchange contributions.
  The underlying cancellation mechanism is
sensitive to both the Higgs- and the gauge-boson sector with
deviations from the SM in particle content or properties immediately
affecting the relevant cross sections. This makes VBS one of
  the most promising classes of processes for the discovery of physics
  beyond the SM in the EW sector. 

In hadronic collisions the scattering of EW gauge bosons can be
accessed in VBS processes which involve the scattering of hadronic
constituents by EW gauge boson exchange. The experimental signature of
such a reaction includes the gauge bosons' decay products and two
distinctive jets resulting from the scattered partons. Because of the
colour-singlet nature of the $t$-channel gauge-boson exchange these
so-called tagging jets tend to be located in the far forward and
backward regions of the detector with a large separation in
rapidity. This feature of VBS reactions helps to identify the signal
in the presence of QCD background processes with large production
rates but rather different signatures. 

VBS processes have received a lot of attention from the 
particle physics community.  Tree-level predictions are available from the dedicated parton-level generator {\tt Phantom}~\cite{Ballestrero:2007xq}, and the multi-purpose tool {\tt Whizard}~\cite{Kilian:2007gr}. 
Next-to-leading-order (NLO) QCD corrections to VBS processes with massive gauge bosons have first been considered in refs.~\cite{Jager:2006zc,Jager:2006cp,Bozzi:2007ur,Jager:2009xx,Denner:2012dz} and been implemented in the framework of the \VBFNLO{} parton-level Monte-Carlo program \cite{Arnold:2008rz}. An alternative option is provided by the multi-purpose program \MGAMCNLO{}~\cite{Alwall:2014hca}. 
The matching of the NLO-QCD calculations with parton-shower (PS) programs according to the \POWHEG~formalism~\cite{Nason:2004rx, Frixione:2007vw} has been considered in refs.~\cite{Jager:2011ms,Jager:2013mu,Jager:2013iza,Jager:2018cyo} and made publicly available in the \POWHEGBOX{}~\cite{Alioli:2010xd} repository. An independent implementation of VBS processes at NLO-QCD accuracy matched with PS in the context of the \HERWIGS{} Monte Carlo program~\cite{Bellm:2015jjp} was presented in ref.~\cite{Rauch:2016upa}. 
More recently, in addition to QCD corrections the NLO electroweak corrections have been considered in refs.~\cite{Biedermann:2016yds,Biedermann:2017bss,Denner:2019tmn,Denner:2020zit,Denner:2021hsa,Denner:2022pwc,Dittmaier:2023nac,Chiesa:2019ulk}.  
For a review of existing work at LO and NLO-QCD accuracy on the representative VBS $\wpwpjj$ channel and a comparison of the individual programs' features we refer the interested reader to  ref.~\cite{Ballestrero:2018anz}.  A LO review of event generators for the VBS $\wzjj$ channel is available from \cite{Proceedings:2018jsb}.

At the CERN Large Hadron Collider (LHC), the ATLAS collaboration reported the observation of EW $\wpmzjj$ production at a center-of-mass energy of $\sqrt{s}=13$~TeV in final states with three identified leptons of electron or muon type in 2018~\cite{ATLAS:2018mxa}. 
The CMS collaboration observed the production of $W^\pm Z$ pairs with leptonic decays in association with two jets at $\sqrt{s}=13$~TeV in 2020, and also provided constraints on anomalous quartic vector boson interactions~\cite{CMS:2020gfh}. In refs.~\cite{ATLAS:2019thr,CMS:2019qfk}, semi-leptonic decay modes were considered. 

In all of these experimental publications, however, the signal simulation was severely limited in accuracy: Refs.~\cite{ATLAS:2018mxa,CMS:2019qfk,CMS:2020gfh} resorted to a leading-order approximation of the signal process. In ref.~\cite{ATLAS:2019thr} NLO-QCD corrections to the on-shell $WZjj$ production process were taken into account, but a factorised ansatz was used to simulate the subsequent decay of the gauge-boson system. This approximation works reasonably well in the resonance region, but fails to provide an accurate description in other regions of phase space, as will be illustrated below. 

In this article we specifically consider the VBS-induced $\wpmzjj$
process. Building on an existing implementation of the VBS $\wzjj$ process~\cite{Jager:2018cyo} in the \POWHEGBOX{}~\cite{Alioli:2010xd} that, however, was limited to fully leptonic decays of the gauge bosons, we provide an updated version of the program accounting for  leptonic, semi-leptonic,  and fully hadronic decay modes of the EW gauge bosons. For each mode, NLO-QCD corrections, off-shell effects, and spin correlations in the decay system are taken into account. Moreover, we provide an option to consider generic extensions of the SM. To illustrate the capabilities of the updated program, phenomenological results are presented for selected scenarios. 

In detail, our work builds on previous developments for VBS-induced $\emmvjj$
and $\pmmvjj$ production in the context of the SM.  In
ref.~\cite{Jager:2018cyo} the NLO-QCD
corrections for these reactions as calculated in \cite{Bozzi:2007ur}
have been implemented in the \POWHEGBOX{}.  Here, we go
beyond this existing implementation by adding semi-leptonic and
hadronic decays of the gauge bosons, and considering an extension of
the SM using the effective field theory (EFT) approach of
ref.~\cite{Degrande:2012wf} that has already been used in the related $\zzjj$ VBS process~\cite{Jager:2013iza}. Effects of dimension-six EFT operators in VBS have also been studied in refs.~\cite{Gomez-Ambrosio:2018pnl,Bellan:2021dcy}.  
In addition, we investigate the relevance of off-shell effects and spin correlations in the decays by an explicit comparison to approximate treatments of these effects.  
We would like to point out the existence of a complementary
calculation~\cite{Denner:2019tmn} for EW $\wpzjj$ calculation in the
fully leptonic decay mode which focuses on perturbative corrections
and provides results including the full NLO QCD and EW
corrections. This calculation has, however, not been matched
  to a parton shower. 

The article is structured as follows: In sec.~\ref{sec:implementation}
we describe the features of the updated implementation of EW $\wzjj$
production in hadronic collisions in the \POWHEGBOX{}. Using this
program, we present some representative numerical results of EW
$\wzjj$ production at the LHC and a potential future circular hadron collider (FCC) operating at an energy of 100~TeV.  We
conclude in sec.~\ref{sec:conclusions}.

%
%=================================================

\section{Details of the \POWHEGBOX{} implementation}
\label{sec:implementation}
In order to provide a Monte-Carlo program for the simulation of EW
$\wzjj$ production in hadronic collisions at NLO+PS accuracy with the
option for various leptonic, semi-leptonic, and hadronic final states
in the context of the SM and a generic model supporting anomalous
interactions in the gauge boson sector we provide appropriate
extensions of the public \POWHEGBOX{} implementation of
Ref.~\cite{Jager:2018cyo}.

We recall that this existing program resorts to tree-level and NLO-QCD
matrix elements for the purely EW processes $pp\to \pmmvjj$ and $pp\to
\emmvjj$ adapted from the \VBFNLO{} parton-level Monte-Carlo
generator~\cite{Arnold:2008rz}.  Even though in the following referred
to as ``EW $\wzjj$ production'' for the sake of brevity, it is
implicitly understood that all resonant and non-resonant diagrams
giving rise to a $\pmmvjj$ or $\emmvjj$ final state system are taken
into account within the so-called VBS approximation.  The VBS
approximation only retains contributions from $t$-channel and
$u$-channel diagrams, but not their interference, and disregards
$s$-channel contributions. When selection cuts typical for an
experimental VBS analysis are applied this approximation has been
found~\cite{Proceedings:2018jsb} to reproduce the full result for the
VBS cross section very well. For instance, for a representative setup
at LO the authors of~\cite{Proceedings:2018jsb} report an agreement at
the level of 0.6\% between calculations based on full matrix elements
and predictions of \VBFNLO{} within the VBS
approximation. 
We note, however, that the validity of this approximation deteriorates when more inclusive selection cuts are applied, see e.g.\ Ref.~\cite{Ballestrero:2018anz} for a comprehensive study of the VBS approximation.  
For this reason we only consider analysis setups with tight VBS cuts in sec.~\ref{sec:pheno}. 
The Cabibbo-Kobayashi-Maskawa matrix is assumed to be diagonal, and
contributions from external top or bottom quarks are not taken into
account. For the updated \POWHEGBOX{} implementation of the EW $\wzjj$ production process presented here we resort to the same approximations.

As long as fully leptonic decays of the gauge bosons are considered,
within the mentioned approximations the structure of the NLO-QCD
corrections does not change if new interactions in the EW gauge-boson
sector are taken into account. The original SM amplitudes for
$\pmmvjj$ and $\emmvjj$ production are structured in a modular way
with leptonic tensors for those building blocks of the relevant
Feynman diagrams that only contain colour-neutral particles, and
hadronic currents accounting for the scattering quarks and, in the
real-emission contributions, gluons. The NLO-QCD corrections only
affect the hadronic currents. Extensions of the SM in the EW sector
thus merely require an appropriate replacement of the leptonic
tensors.

We provide such an extension in the framework of the effective field
theory (EFT) approach of Ref.~\cite{Degrande:2012wf} accounting for
anomalous interactions in the EW gauge boson sector by an extension of
the SM Lagrangian with operators of higher mass dimension,
\beq
\label{eq:L_EFT}
\mc{L}_\mr{eff} = 
\mc{L}_\mr{SM}  + 
\sum_{d>4}\sum_i \frac{c_i^{(d)}}{\Lambda^{d-4}}\mc{O}_i^{(d)}   \,. 
\eeq
Here, $d$ denotes the mass dimension of the operators
$\mc{O}_i^{(d)}$, and the $c_i^{(d)}$ are the expansion coefficients. The sum
over $d$ includes contributions of all higher-dimensional operators
starting from $d=6$. The summation index $i$ runs over all
non-vanishing operators of a given mass dimension. The parameter
$\Lambda$ denotes the energy scale up to which the EFT is supposed to
be valid. We assume $\Lambda$ to be much larger than the EW scale and
thus restrict ourselves to the contribution of operators up to
dimension six.

Following Ref.~\cite{Degrande:2012wf} we consider three independent
operators that conserve charge ($C$) and parity ($P$),
\begin{align}
\mathcal{O}_{WWW} &= \mathrm{Tr}[\hat{W}_{\mu\nu}\hat{W}^{\nu\rho}\hat{W}_{\rho}^{\mu}], \label{eq:Cwww}\\
\mathcal{O}_{W} &= (D_{\mu}\Phi)^{\dag}\hat{W}^{\mu\nu}(D_{\nu}\Phi), \label{eq:Cw}\\
\mathcal{O}_B &= (D_{\mu}\Phi)^{\dag}\hat{B}^{\mu\nu}(D_{\nu}\Phi),\label{eq:Cb}
\end{align}
and two C and/or P violating operators,  
\begin{align}
\mathcal{O}_{\tilde{W}WW}  &=  \mathrm{Tr}[\tilde{W}_{\mu\nu}\hat{W}^{\nu\rho}\hat{W}_{\rho}^{\mu}], \label{eq:Cpwww}\\
\mathcal{O}_{\tilde{W}} &= (D_{\mu}\Phi)^{\dag}\tilde{W}^{\mu\nu}(D_{\nu}\Phi).\label{eq:Cpw}
\end{align}
These operators are constructed from the Higgs doublet field $\Phi$
and the electroweak field strength tensors $W_{\mu\nu}^{a}$
$(a=1,2,3)$ and $B_{\mu\nu}$,
\begin{align}
W^{a}_{\mu\nu}&=\partial_{\mu}W^{a}_{\nu}-\partial_{\nu}W^{a}_{\mu}-g\epsilon_{abc}W_{\mu}^{b}W_{\nu}^{c} ,\label{eq:2.10}\\ 
B_{\mu\nu}&=\partial_{\mu}B_{\nu}-\partial_{\nu}B_{\mu}\label{eq:2.11},
\end{align}
with the $U(1)$ and $SU(2)$ gauge fields $B_{\mu}$ and $W^a_{\mu}$ and
their respective couplings $g'$ and $g$. The $\sigma^a$ denote the
Pauli matrices.
The covariant derivative $D_\mu$ is given by
\begin{equation}
D_{\mu} = \partial_{\mu}+igW_{\mu}^{a}\dfrac{\sigma^{a}}{2}+\dfrac{1}{2}i g'B_{\mu},
\end{equation}
and the modified field strength tensors $\hat{W}_{\mu\nu}$ and $\hat{B}_{\mu\nu}$ are defined by \begin{align}
\hat{W}_{\mu\nu} &= i g \dfrac{\sigma^{a}}{2} W_{\mu\nu}^{a}, \label{Wmunu}\\
\hat{B}_{\mu\nu} &=\dfrac{i}{2}g' B_{\mu\nu}, \label{Bmunu}\\
\left[D_{\mu},D_{\nu}\right] &=\hat{W}_{\mu\nu}+\hat{B}_{\mu\nu}\,,
\end{align}
while the modified dual field strength tensor is given by
\begin{equation}\label{Wtilde}
\tilde{W}_{\mu\nu} = \epsilon_{\alpha\beta\mu\nu}\hat{W}^{\alpha\beta}.
\end{equation}
To simplify our notation we denote the coefficients of the operators
of eqs.~(\ref{eq:Cwww})--(\ref{eq:Cpw}) that appear in the EFT
expansion of eq.~(\ref{eq:L_EFT}) up to dimension six as
\begin{equation}\label{C_i first}
C_i \equiv  \dfrac{c_i^{(6)}}{\Lambda^2}. 
\end{equation}

In the following, instead of a numbered index $i$ we use the label of
the corresponding operator to identify each operator coefficient. For
instance, $C_{WWW}$ is the properly normalized coefficient of the
$\mathcal{O}_{WWW}$ operator.

In the actual calculation of scattering cross sections care has to be
taken to ensure a consistent EFT expansion up to the desired order in
$1/\Lambda^2$. Schematically, the operators of different mass
dimension enter in the relevant matrix elements squared as
\begin{align}\label{eq:mat-exp}
 \vert \mathcal{M}_{SM} + \mathcal{M}_{dim-6} +\mathcal{M}_{dim-8} + \cdots \vert^2 &= \underbrace{\vert \mathcal{M}_{SM} \vert^2}_{\Lambda^0} + 
 \underbrace{2 Re(\mathcal{M}_{SM}\mathcal{M}_{dim-6}^\star)}_{\Lambda^{-2}} \nonumber \\ 
& + \underbrace{\vert\mathcal{M}_{dim-6}\vert^2+2 Re(\mathcal{M}_{SM}\mathcal{M}_{dim-8}^\star)}_{\Lambda^{-4}}+\cdots\,.
\end{align}   
Thus, if one truncates the EFT expansion at order $1/\Lambda^2$, in
addition to the pure SM contribution one should only keep the
interference term $2 Re(\mathcal{M}_{SM}\mathcal{M}_{dim6}^\star)$,
and disregard the quadratic term $ \vert\mathcal{M}_{dim6}\vert^2$,
which is part of the $\mc{O}(1/\Lambda^4)$ result.
However, in the past in many applications this quadratic term was
considered as part of the ``dimension six'' results. Below we
therefore consider both options and refer to them as \smlin{} and
\smquad{}. 

\begin{figure}[t]
\begin{center}
\subfloat[][]{
\includegraphics[width=0.35\textwidth]{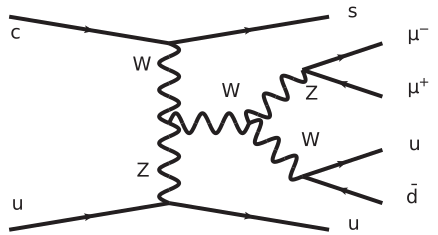}}
\subfloat[][]{
\includegraphics[width=0.3\textwidth]{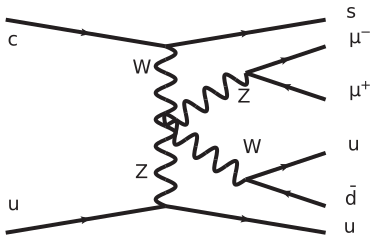}}
\subfloat[][]{
\includegraphics[width=0.35\textwidth]{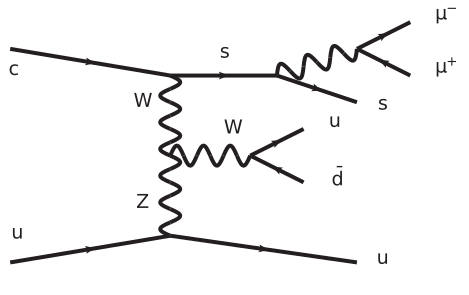}}
\qquad
\subfloat[][]{
\includegraphics[width=0.325\textwidth]{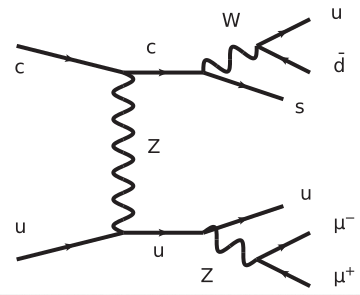}}
\subfloat[][]{
\includegraphics[width=0.35\textwidth]{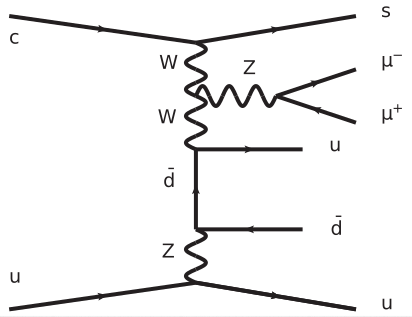}}
\subfloat[][]{
\includegraphics[width=0.325\textwidth]{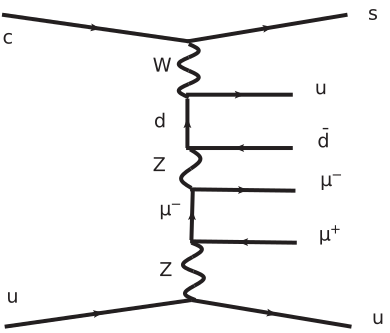}}
\caption{\label{fig:diagrams}
Representative Feynman diagrams for the VBS-process $ u \,\, c \rightarrow u \,\, s \,\, \bar{d} \,\, u \,\, \mu^{+} \,\, \mu^{-} $: 
(a),(b) genuine vector-boson scattering diagrams, 
(c),(d) diagrams including gauge-boson emission from a quark line, 
(e) singly-resonant, and (f) non-resonant diagrams. 
}
\end{center}
\end{figure}

In addition to the $\pmmvjj$ and $\emmvjj$ final states provided in
ref.~\cite{Jager:2018cyo}, in this work we also
implemented 
EW production of a $\vevv$, a $\qqmm$, a $\emQQ$, a $\pmQQ$, or a $\qqQQ$
system, respectively, in association with two tagging jets (here $q$
and $Q$ refer to massless quarks of different types). 
Representative diagrams for the partonic channel $u c \to u s\, \bar d u\, \mu^+\mu^-$ are shown in fig.~\ref{fig:diagrams}. 
%
%%%%%%%%
%%%%%%%%%
For each channel,  not only resonant diagrams related to the leptonic or hadronic decay
of a $W$ or $Z$ boson are taken into account, but also non-resonant
diagrams resulting in the same final state.  For simplicity we will
refer to the previously listed processes (including off-resonant
contributions) as fully leptonic, leptonic-invisible, semi-leptonic,
and fully hadronic decay modes.
In the case of semi-leptonic and fully hadronic decay modes we do not
take QCD corrections to the decays into account, and we neglect QCD
corrections connecting the $\wzjj$ production with the decay part of
the considered final state. The latter type of corrections are
expected to be negligible.  Corrections to the hadronic decays of the
$Z$ and $W$ bosons are accounted for by the multi-purpose Monte-Carlo
programs matched to our NLO-QCD calculation.

%%%%%%%%%%%% 
For each final state, at Born level singularities in the production
cross section arise from diagrams with the $t$-channel exchange of a
photon of very low virtuality $Q^2$. Such contributions are entirely
negligible after selection cuts on the tagging jets are applied and
can thus be removed already at generation level by a cut, \beq
Q^2_\mr{min}= 4~\text{GeV}^2\,.  \eeq
To further improve the numerical efficiency of the Monte-Carlo
integration additionally a Born-suppression factor of the form
\beq
F(\Phi) =
\left(\frac{p_{T,1}^2}{p_{T,1}^2+\Lambda_\Phi^2}\right)^2
\left(\frac{p_{T,2}^2}{p_{T,2}^2+\Lambda_\Phi^2}\right)^2\,.
\eeq
can be employed. Here, the $p_{T,i}$ denote the transverse momenta of
the final-state partons of the underlying Born configuration $\Phi$,
and $\Lambda_\Phi$ is a technical parameter, by default set to 10~GeV.

We remind the reader that contributions with a pair of
same-type charged fermions ($f$) in the final state
cannot only stem from decays of the $WZ$ system, but also from
diagrams where a photon decays into a fermion pair. An additional type
of singularity at Born level arises from diagrams where such a photon
exhibits very low virtuality. For analyses that require a fermion pair
with an invariant mass close to the mass of the $Z$~boson, such
singularities can easily be removed already at generation level by an
invariant mass cut on the respective lepton pair.  The requirement
\beq
m_{f \bar f}>0.5~\text{GeV}
\eeq
suffices to remove any potentially problematic contributions from
photons of very low virtuality at generation level.

\section{Phenomenological results}
\label{sec:pheno}
In the following we will provide some phenomenological results
generated by our implementation in version 2 of the \POWHEGBOX. 
For all of these results
we use the following general settings: We set the Fermi constant to $
G_{\mu} = 1.1663787 \cdot 10^{-5} \, \mathrm{GeV^{-2}} $.  For the
masses and widths of the EW bosons we use: ${m_H = 125.25 \,
  \mathrm{GeV}}$, ${m_W = 80.377 \, \mathrm{GeV}}$, ${m_Z = 91.1876 \,
  \mathrm{GeV}}$, ${\Gamma_H = 0.0032 \, \mathrm{GeV}}$, ${\Gamma_W =
  2.085 \, \mathrm{GeV}}$, and ${\Gamma_Z = 2.4952 \, \mathrm{GeV}}$.
The EW coupling, $\alpha_{em}$, is calculated therefrom via tree-level
EW relations.

In addition to the \POWHEGBOX{} we also use the tool-chain
\texttt{MadGraph5\_aMC@NLO}~\cite{Alwall:2014hca,Frederix:2018nkq}
including the program \texttt{MadSpin}~\cite{Artoisenet:2012st} for
comparison to the \POWHEGBOX{} results.  To simulate the parton shower
we use \PYTHIAE{}, version 8.245, with the {\tt Monash2013}
tune~\cite{Skands:2014pea}. Hadronisation, MPI, and QED emissions are
turned off in order to isolate the effect of the shower and
matching. We note that in realistic simulations non-perturbative
effects have to be considered~\cite{Bittrich:2021ztq}. For
reconstructing jets we resort to
\texttt{FastJet}~\cite{Cacciari:2011ma}, version 3.3.4. For all
results presented in this section jets are clustered via the
anti-$k_T$ algorithm~\cite{Cacciari:2008gp} with a radius-parameter of
$R=0.4$, and the \texttt{NNPDF31\_nlo\_as\_0118} (ID 303400)
set~\cite{NNPDF:2017mvq} of parton distribution functions (PDFs) is
used as provided by version~6.3.0 of the {\tt LHAPDF}
library~\cite{Buckley:2014ana}.

%%%%%%%%%%%%%%%%
\subsection{EFT results}
In this subsection we explore the impact of SM extensions in the EFT
framework introduced in sec.~\ref{sec:implementation}. We individually
set the coefficient of each EFT operator defined in
eqs.~(\ref{eq:Cwww})-(\ref{eq:Cpw}) to the largest value compatible
with the experimental limits of ref.~\cite{CMS:2021icx} while setting the
coefficients of all other EFT operators to zero. As it turned out that
the impact of the $\mc{O}_{WWW}$ operator is most pronounced, below we
only display results obtained for non-vanishing values of the
$C_{WWW}$ coefficient. For instance, results obtained with
non-vanishing values of the $C_{W}$ operator coefficient are basically
identical to the SM results and will thus not be further discussed
here.

For the unitarisation of our EFT predictions we
proceeded along the lines of ref.~\cite{Sirunyan_2020} where unitarity
violations are avoided by using appropriate cuts on the invariant mass
of the vector bosons produced in a VBS reaction.  We calculated the
limits beyond which unitarity violations are to be expected for the 
setups considered in this work using the tool \texttt{calc-formfactor}~\cite{Feigl,Schlimpert}
%version~1.4.0,  
that is available within the \VBFNLO{}
package~\cite{Baglio:2011juf,Baglio:2014uba}. We found that
unitarity violations would occur only beyond scales relevant for the results shown below. 

Throughout this subsection we use a renormalisation scale, $\mur=\xir  \mu_0$, and factorisation scale, $\muf=\xif  \mu_0$, that is expected to optimally account for
the region of high transverse momenta where the effects of the
dimension-six operators are expected to have the largest
impact~(c.f.\ ref.~\cite{Jager:2013iza}). 
The scale $\mu_0$ is given by
\begin{equation}\label{eq:scales}
\mu_0 = \dfrac{1}{2}\left(E_{T,W}+E_{T,Z}+ \sum_f^{\mr{n_{part}}} p_{T,f}\right)\,, 
\end{equation}
where the sum includes the transverse momenta $p_{T,f}$ of all
$\mr{n_{part}}$ final-state partons of a considered Born-type or
real-emission configuration. In addition we define
\begin{equation}\label{Eq:ETW neu}
E_{T,W} = \sqrt{m_W^2+p_{T,W}^2}, \qquad E_{T,Z} = \sqrt{m_Z^2+p_{T,Z}^2},
\end{equation}
where $p_{T,Z}$ and $p_{T,W}$ are the transverse momenta of the muon
pair and the positron-neutrino pair of the fixed-order configuration,
respectively.
The factors $\xir$ and $\xif$ are varied between $0.5$ and $2$ with 7-point variation in our NLO+PS simulations and set to one for the fixed-order calculations.

For our representative numerical studies we use settings inspired by
the ATLAS analysis described in ref.~\cite{ATLAS:2016bkj}.  In
particular, we construct jets using the anti-$k_T$ algorithm with
$R=0.4$. In the following we will denote the jets with index $j_1$ for the hardest jet and $j_2,j_3,...$ for further jets ordered by their transverse momentum. The two hardest jets are identified as tagging jets and are
required to have a transverse momentum, rapidity and invariant mass of
\begin{equation}
\label{eq:jet-cuts1}
p_{T,j}^\mr{tag} > 30 \,  \mathrm{GeV}, \qquad \vert y_j^\mr{tag} \vert < 4.5, \qquad m_{jj}^\mr{tag} > 500 \, \mathrm{GeV}.
\end{equation}
Moreover, we only keep events where the tagging jets lie in opposite hemispheres, 
\begin{equation}
\label{eq:jet-cuts2}
\eta_{j_1}^\mr{tag}\cdot \eta_{j_2}^\mr{tag} < 0,
\end{equation}
and have a large rapidity separation of 
\begin{equation}
\label{eq:jet-cuts3}
\Delta y_{jj}^\mr{tag} = \vert y_{j_1}^\mr{tag}-y_{j_2}^\mr{tag} \vert \geq 2.5. 
\end{equation}
  
To identify additional non-tagging jets we require them to be located in a rapidity range of 
\begin{equation}
\vert y_j \vert < 4.5\,.
\end{equation}
No other cuts are applied on non-tagging jets unless specifically stated otherwise.

\begin{figure}[t!]
\centering
\subfloat[][]{
\includegraphics[width=0.5\textwidth]{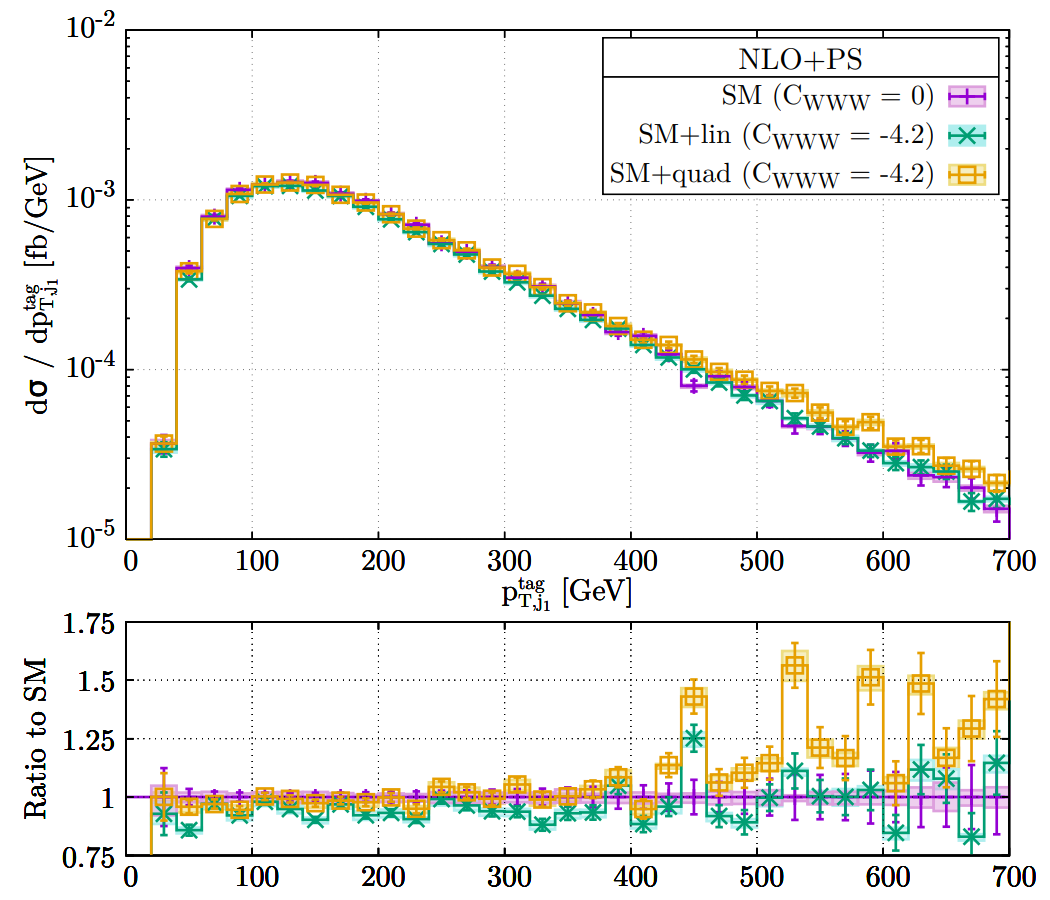}}
\subfloat[][]{
\includegraphics[width=0.5\textwidth]{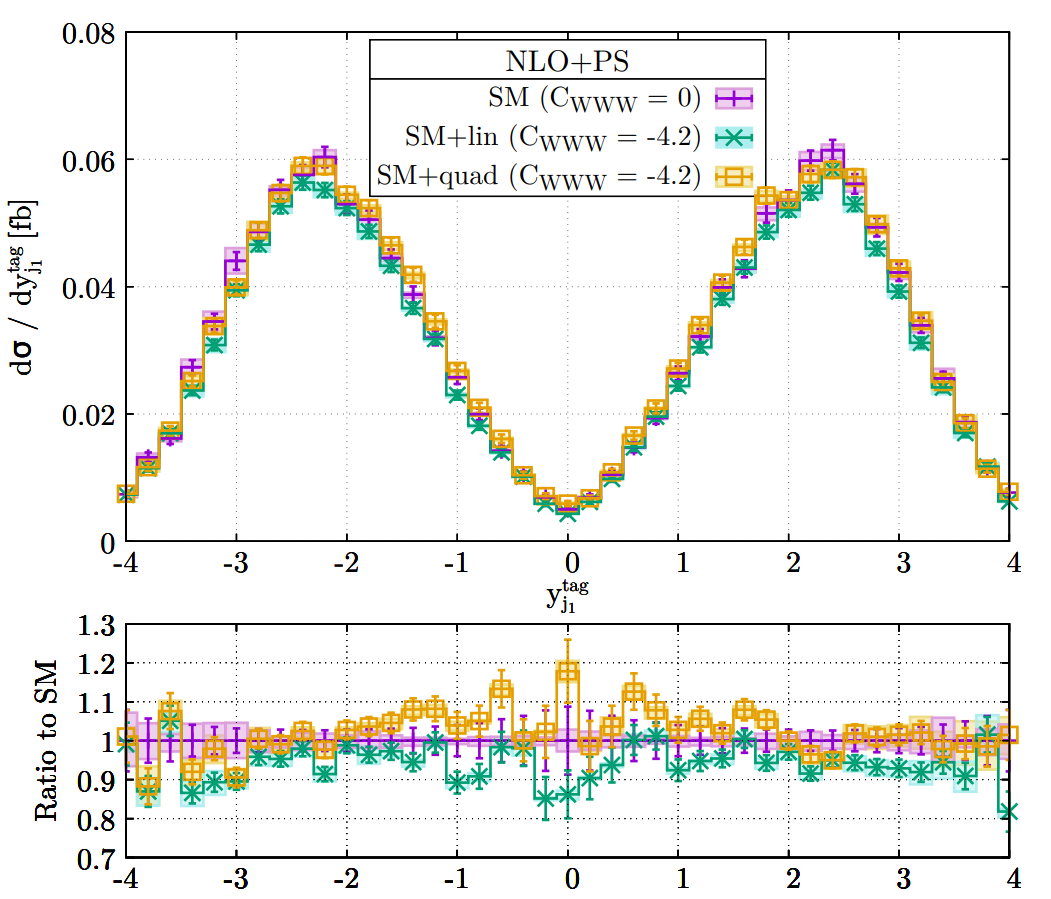}}
\qquad
\subfloat[][]{
\includegraphics[width=0.5\textwidth]{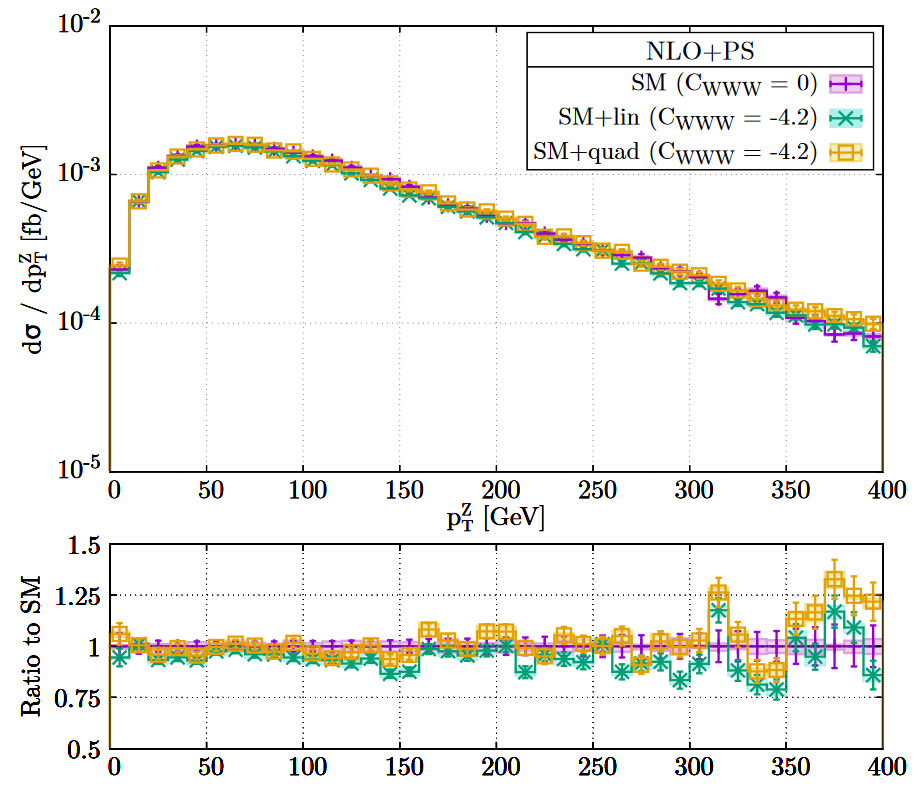}}
\subfloat[][]{
\includegraphics[width=0.5\textwidth]{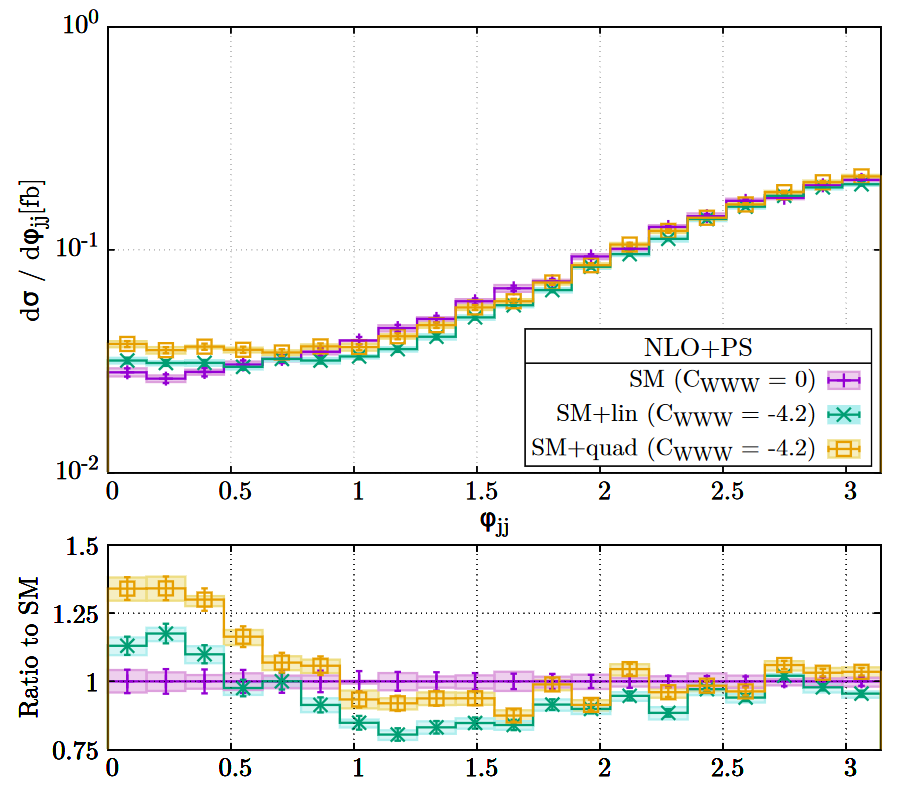}}
\caption{\label{fig:NP_1} NLO+PS predictions for $pp\to \pmmvjj$ at
  the LHC with $\sqrt{s}=13$~TeV within the cuts of
  eqs.~(\ref{eq:jet-cuts1})--(\ref{eq:gap-cut}) for the
  \smlin{}~(green) and the \smquad{} case (orange) with $C_{WWW} =
  -4.2 \, \mathrm{TeV}^{-2}$, and within the SM
(purple).  The upper panels show the transverse momentum of the
  hardest tagging jet (a), the rapidity of the hardest tagging jet
  (b), the reconstructed transverse momentum of the $Z$ boson (c), and
  the azimuthal angle separation of the tagging jets (d). The
  respective lower panels show the ratios of the \smlin{} and
  \smquad{} predictions to the pure SM results.  }
\end{figure}

\begin{figure}[t!]
\centering
\subfloat[][]{
\includegraphics[width=0.5\textwidth]{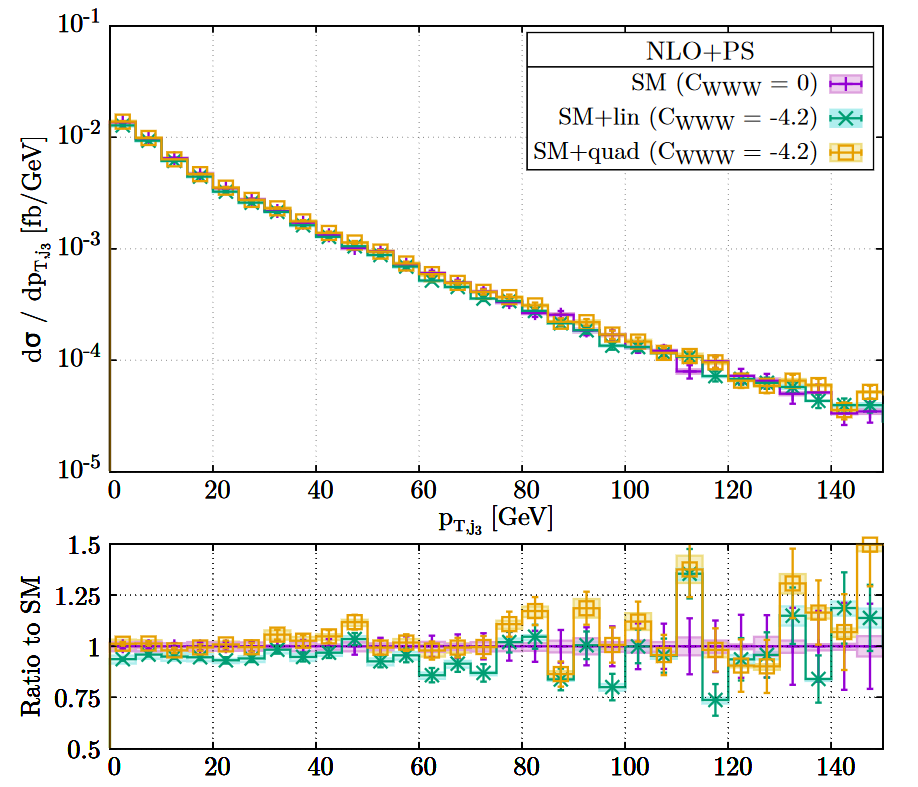}}
\subfloat[][]{
\includegraphics[width=0.5\textwidth]{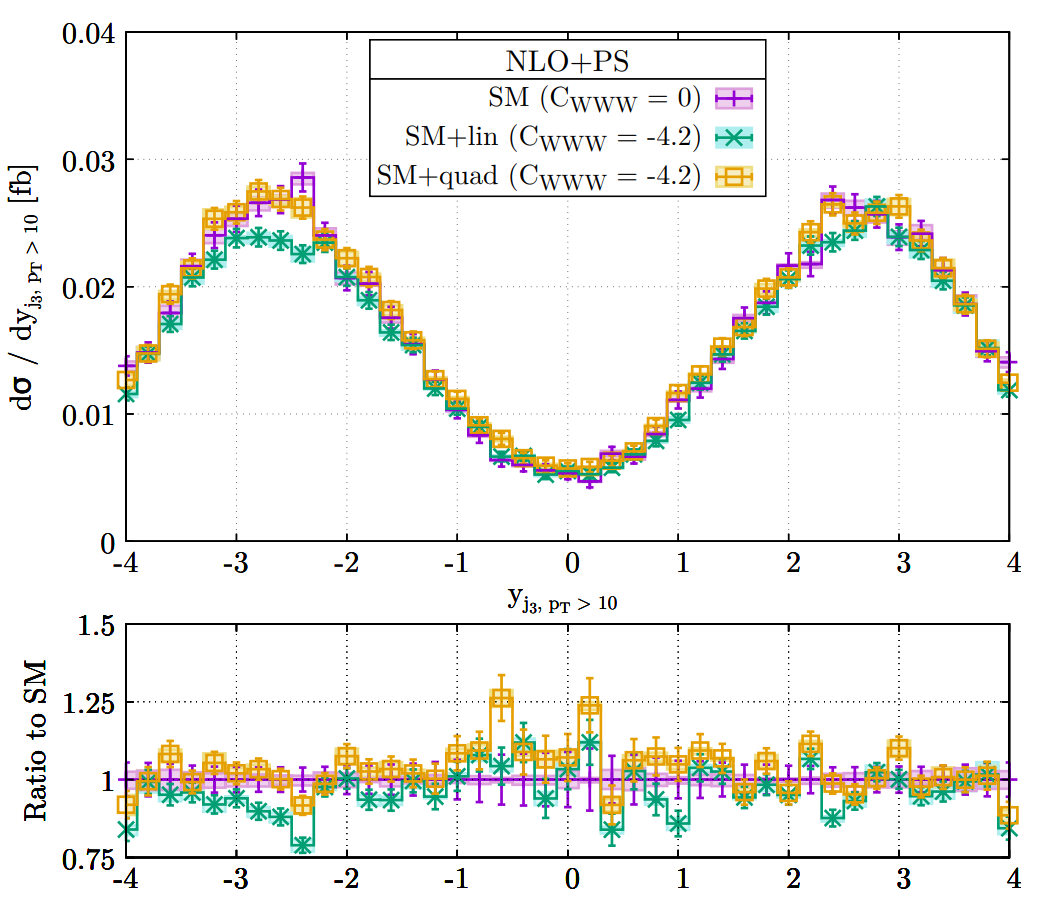}}
\caption{\label{fig:NP_1.5} NLO+PS predictions for $pp\to \pmmvjj$ at
  the LHC with $\sqrt{s}=13$~TeV within the cuts of
  eqs.~(\ref{eq:jet-cuts1})--(\ref{eq:gap-cut}) for the
  \smlin{}~(green) and the \smquad{} case (orange) with $C_{WWW} =
  -4.2 \, \mathrm{TeV}^{-2}$, and within the SM
(purple).  The upper panels show the transverse momentum of the 3rd jet (a) and the rapidity of the 3rd jet for a $p_T$ cut of $p_{T,j_3} > 10 \, \mathrm{GeV}$
  (b). The
  respective lower panels show the ratios of the \smlin{} and
  \smquad{} predictions to the pure SM results.  }
\end{figure}

In our fixed-order results the final state contains two muons, one
positron and one neutrino. Within the \PYTHIA{} setup we consider,
i.e.\ in the absence of QED radiation in the parton shower and without
hadron decays, the events we simulate at NLO+PS level do not exhibit
any additional leptons or neutrinos.
For the charged leptons $\ell$ we demand
\begin{equation}
\label{eq:lep-cuts1}
p_{T,\ell} > 15 \,  \mathrm{GeV}, \qquad \vert y_\ell \vert < 2.5.
\end{equation}
We do not apply any cuts on the neutrino.  Furthermore, we require a
clear separation of the tagging jets and the charged leptons, i.e.\ we require a separation
in the rapidity-azimuthal angle plane of \beq
\label{eq:jl-cut}
R_{j\ell} > 0.3\,.
\eeq 
For the muons which are stemming from the $Z$ decay we additionally demand 
\beq
\label{eq:cut-rll}
R_{\mu\mu} > 0.3\,, \eeq as well as that their reconstructed invariant
mass, $m_\mr{inv}^Z$, lies in a window around the physical $Z$-boson
mass of
\begin{equation}
\label{eq:mz-cut}
66 \, \mathrm{GeV} < m_\mr{inv}^Z < 116 \, \mathrm{GeV}\,. 
\end{equation}
Finally, we also require all charged leptons to lie in the rapidity gap between the two tagging jets 
\begin{equation}
\label{eq:gap-cut}
\mathrm{min}(y_{j_1}^\mr{tag},y_{j_2}^\mr{tag}) < y_\ell < \mathrm{max}(y_{j_1}^\mr{tag},y_{j_2}^\mr{tag}).
\end{equation}

Let us now discuss results for the LHC with a center-of-mass energy of
$\sqrt{s}=13 \, \mathrm{TeV}$.  In fig.~\ref{fig:NP_1} and fig.~\ref{fig:NP_1.5} we compare SM
results at NLO+PS accuracy for selected distributions of the tagging
jets and the leptons with those obtained in the EFT framework of
sec.~\ref{sec:implementation}. In particular, we set the operator
coefficient $C_{WWW}$ to the maximal negative value compatible with current
experimental limits, i.e.\ $C_{WWW}= -4.2 \, \mathrm{TeV}^{-2}$, and
consider separately the case where only the linear term of the EFT
expansion sketched in eq.~(\ref{eq:mat-exp}) is taken into account
(dubbed \smlin), and the case where additionally the quadratic term is retained (referred to as \smquad).  
%%%%%%%
%

%
%%%%%
%
For the \smlin{} implementation we find only small differences to the
SM results for all considered distributions. These effects are best
visible for the azimuthal angle separation of the two tagging jets,
$\Phi_{jj}$. In particular for $\Phi_{jj}\lesssim \pi/2$ the shape is
slightly different from the SM case. Larger differences to the SM case
are found for the \smquad{} implementation. However, we would like to
remind the reader that the limits chosen for $C_{WWW}$ have been
derived for the \smlin{} implementation and that the \smquad{}
version is shown only for the purpose of
comparison. The \smlin{} and \smquad{} predictions can barely be
distinguished from the respective SM results for the transverse
momentum and the rapidity distributions of the hardest tagging jet and
the transverse momentum of the $Z$~boson. The latter is reconstructed
from the momenta of the muon pair closest in invariant mass to $m_Z$.

\begin{figure}[t!]
\centering
\subfloat[][]{
\includegraphics[width=0.5\textwidth]{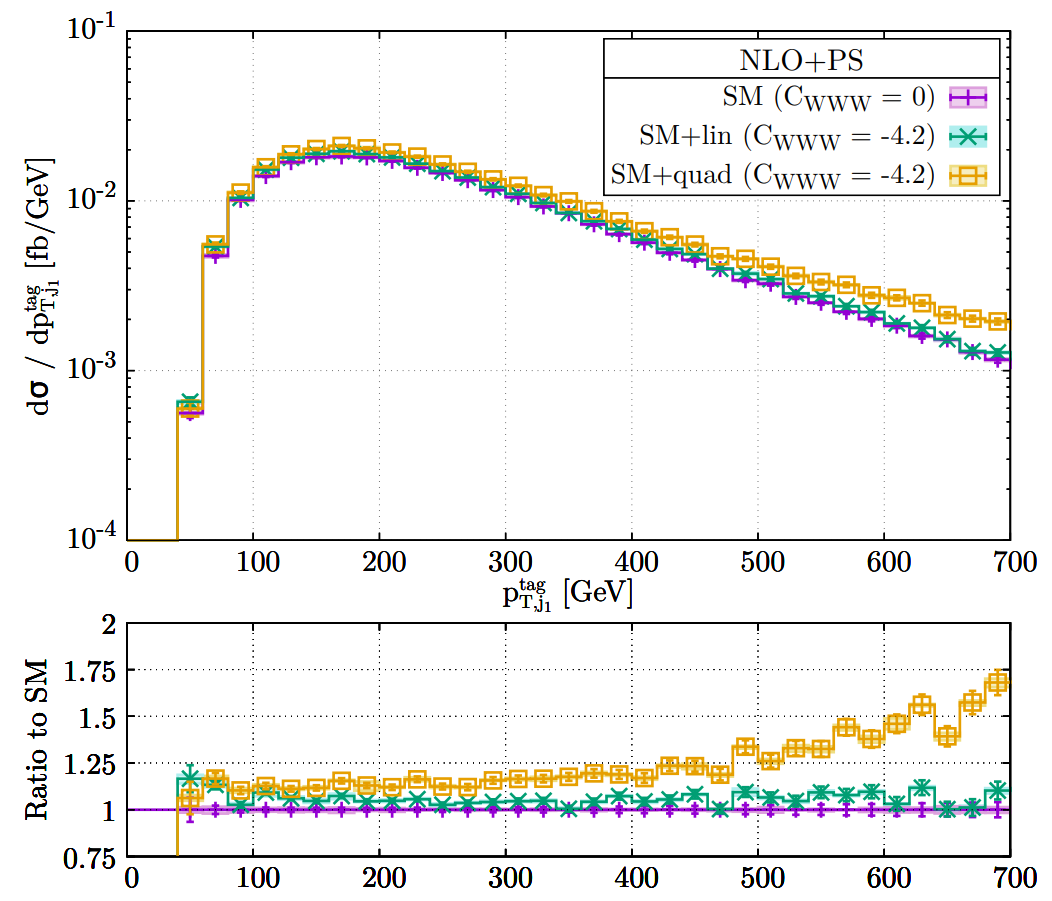}}
\subfloat[][]{
\includegraphics[width=0.5\textwidth]{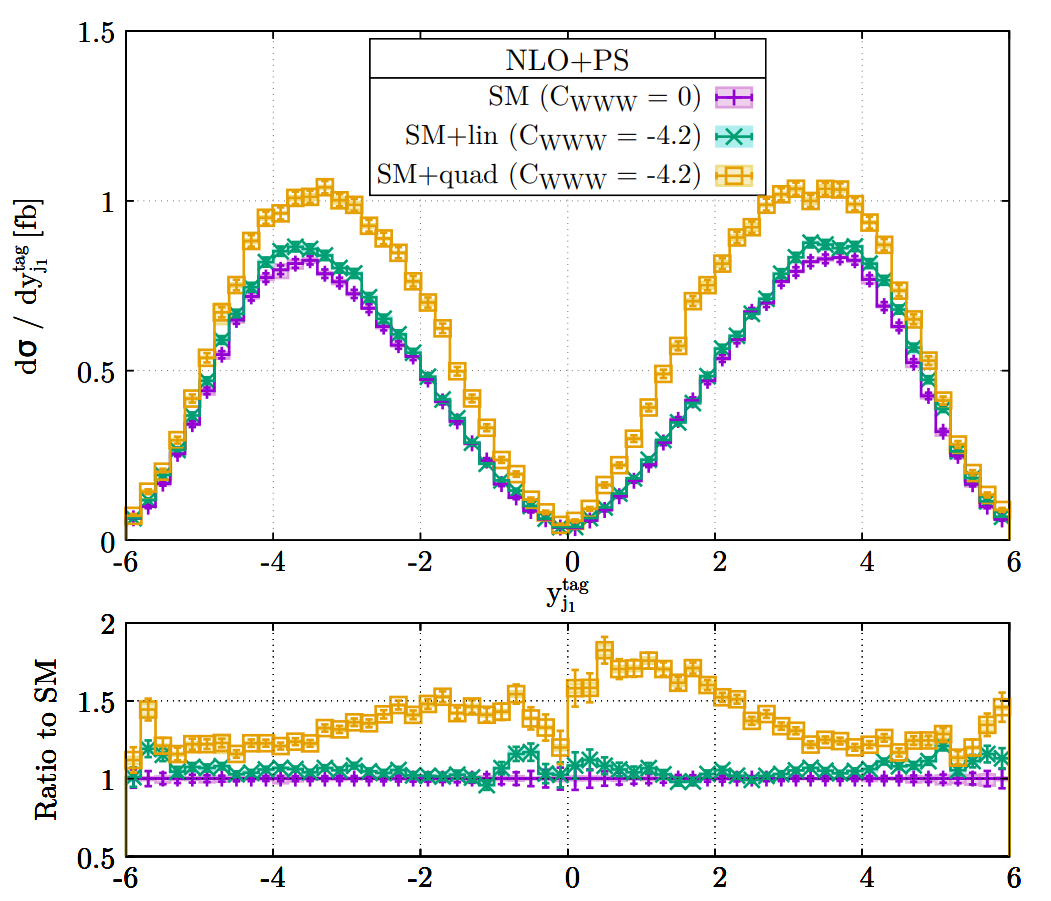}}
\qquad
\subfloat[][]{
\includegraphics[width=0.5\textwidth]{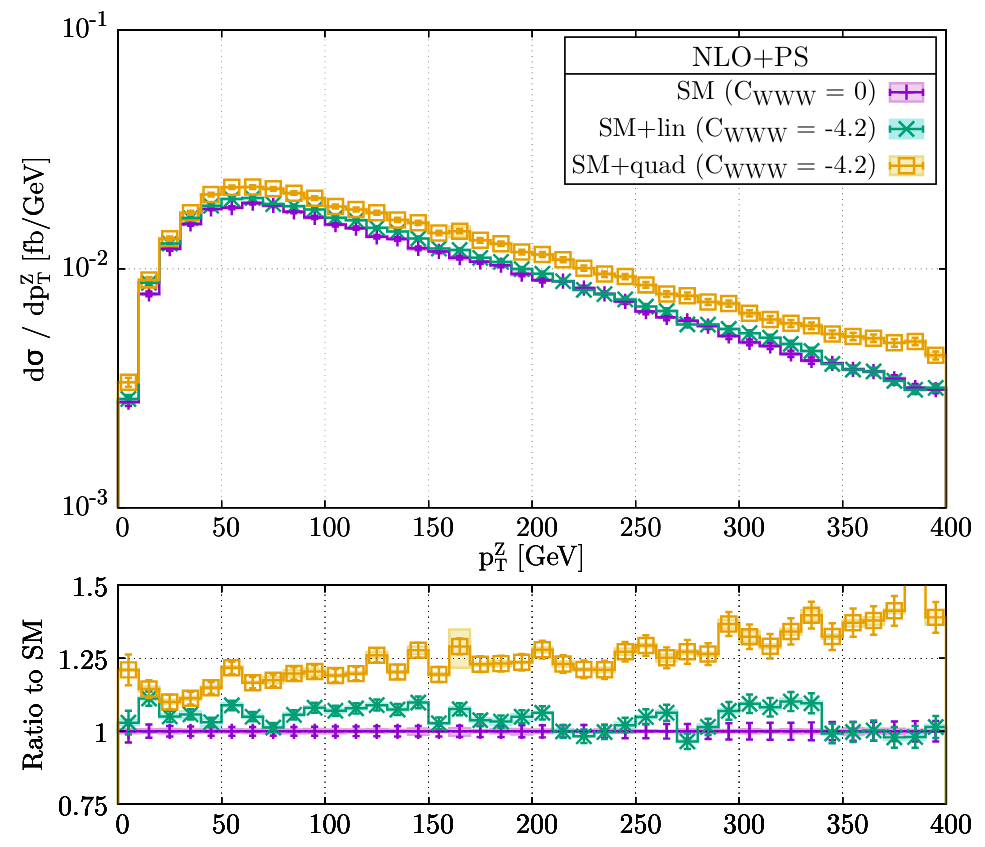}}
\subfloat[][]{
\includegraphics[width=0.5\textwidth]{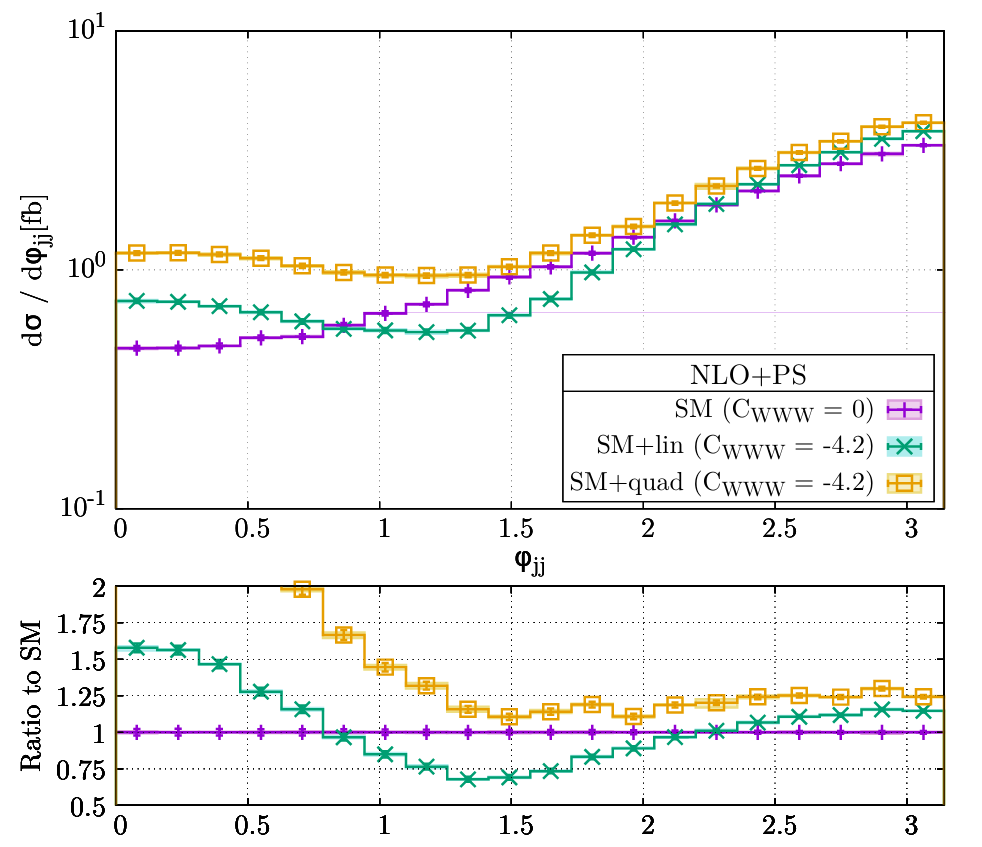}}
\caption{\label{fig:NP_2}
Same as in fig.~\ref{fig:NP_1}, but for the FCC with an energy of $\sqrt{s}=100$~TeV  and with the cuts of eqs.~(\ref{eq:cut-rll})--(\ref{eq:jl-cut_FCC}).
}
\end{figure}

\begin{figure}[t!]
\centering
\subfloat[][]{
\includegraphics[width=0.5\textwidth]{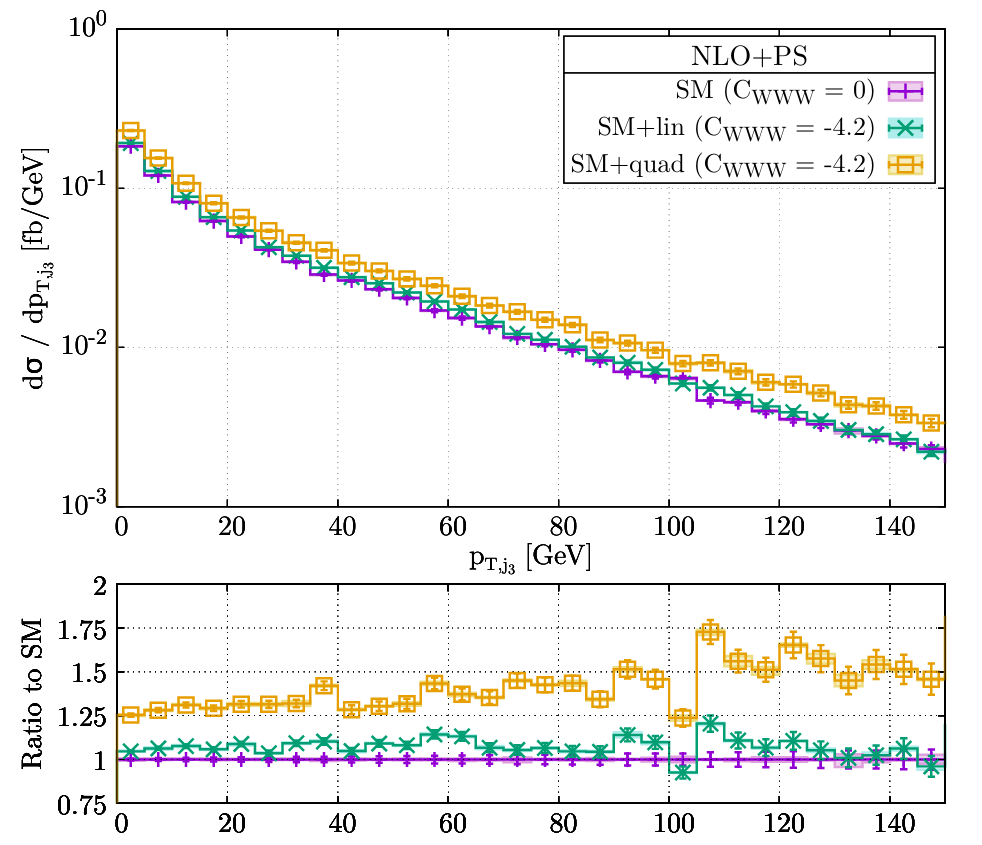}}
\subfloat[][]{
\includegraphics[width=0.5\textwidth]{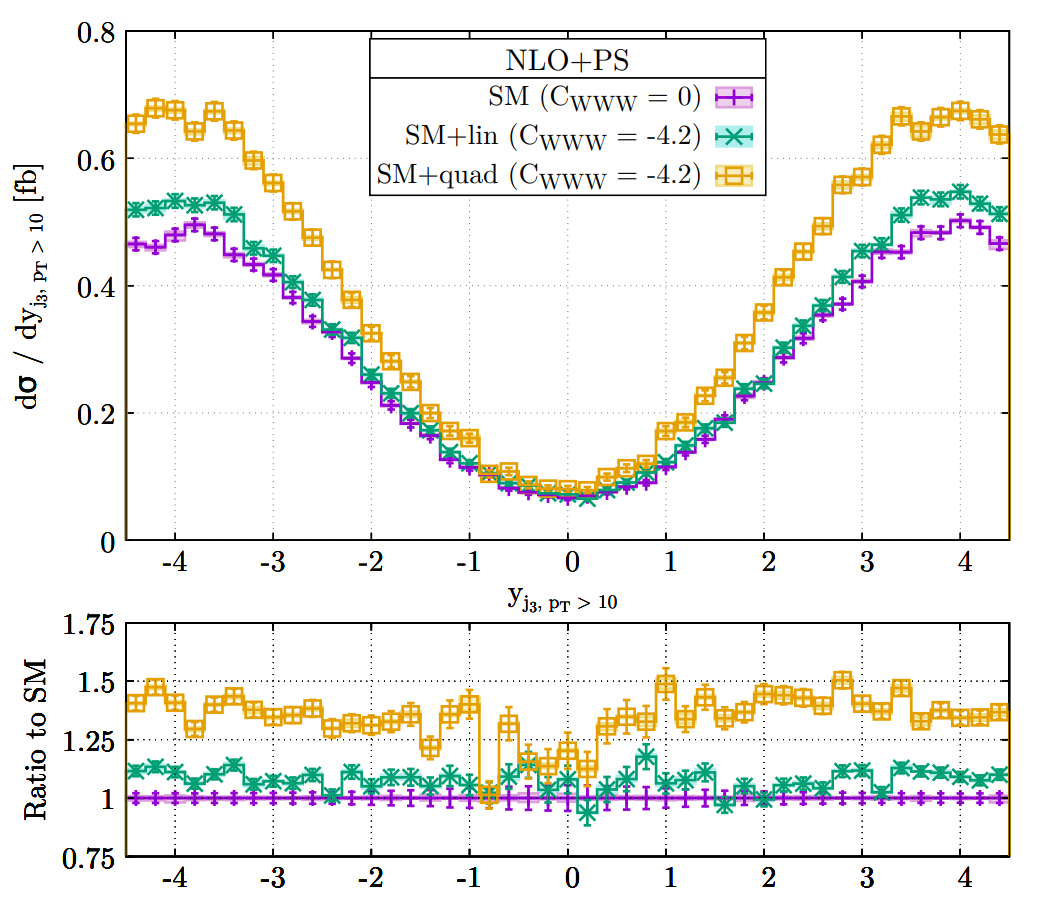}}
\caption{\label{fig:NP_2.5} Same as in fig.~\ref{fig:NP_1.5}, but for the FCC with an energy of $\sqrt{s}=100$~TeV and with the cuts of  eqs.~(\ref{eq:cut-rll})--(\ref{eq:jl-cut_FCC}).}
\end{figure}

In fig.~\ref{fig:NP_2} and fig.~\ref{fig:NP_2.5} we display the same observables as in
fig.~\ref{fig:NP_1} and fig.~\ref{fig:NP_1.5} for a potential future circular collider (FCC)
with proton-proton collisions at a center-of-mass energy of $\sqrt{s}=100~\mathrm{TeV}$.
For the FCC discussion we use the following setup inspired by ref.~\cite{Jager:2017owh}.
We use the same renormalisation and factorisation scale as in eq.~(\ref{eq:scales}), but apply stronger cuts on the tagging jets.
More precisely, we require 
\begin{equation}
\label{eq:jet-cuts1_FCC}
p_{T,j}^\mr{tag} > 50 \,  \mathrm{GeV}, \qquad \vert y_j^\mr{tag} \vert < 6, \qquad m_{jj}^\mr{tag} > 2500 \, \mathrm{GeV}.
\end{equation}
Additionally, the tagging jets have to fulfill 
\begin{equation}
\label{eq:jet-cuts2_FCC}
\eta_{j_1}^\mr{tag}\cdot \eta_{j_2}^\mr{tag} < 0, 
\quad 
\Delta y_{jj}^\mr{tag} = \vert y_{j_1}^\mr{tag}-y_{j_2}^\mr{tag} \vert \geq 5. 
\end{equation} 
For the charged leptons we require
\begin{equation}
\label{eq:lep-cuts1_FCC}
p_{T,\ell} > 20 \,  \mathrm{GeV}, \qquad \vert y_\ell \vert < 5,
\end{equation}
as well as a separation of the tagging jets and the charged leptons in the rapidity-azimuthal angle plane of \beq
\label{eq:jl-cut_FCC}
R_{j\ell} > 0.4\,.
\eeq 
We keep the requirements of eqs.~(\ref{eq:cut-rll})--(\ref{eq:gap-cut}).
%
%%%%%%%%
%

%
%%%%%%%
%
Differences between the SM and the \smquad{} implementation are
clearly enhanced at this energy for all distributions. However, the
azimuthal angle separation of the tagging jets exhibits sensitivity to
the effects of the \smlin{} implementation.

%%%%%%%%%%

\subsection{Leptonic decays}

The objective of this subsection is to explore the relevance of
simulating the full leptonic final state of VBS-induced $\emmvjj$
production as opposed to approximations where a $W^+Z$ boson pair is
produced on-shell and combined with a simulation for the decays of
these bosons into the desired leptonic final state. 
Such an approximation is used, for instance, in the search for anomalous EW production of vector boson pairs in association with two jets by the CMS collaboration~\cite{CMS:2019qfk} and in the search for EW diboson production in association with a high-mass dijet system in semi-leptonic final states by the ATLAS collaboration~\cite{ATLAS:2019thr}.
Since QCD
corrections do not directly affect the leptonic decays, we conduct
this discussion at LO.
To that end we compare our \POWHEGBOX{} implementation for
$\emmvjj$ production via VBS with two alternative ones: The first is
using \MGAMCNLO \, where a $W^+Zjj$ final state is
produced on-shell.
The decays of the two bosons are afterwards simulated via
\texttt{MadSpin} (this simulation is denoted by
\texttt{MG5+MadSpin} below). The other implementation is using the LO version of
\MGAMCNLO{} and includes off-shell contributions and spin-correlations
in the lepton system (this implementation will be denoted by
\texttt{MG5-full} in the following). 
This comparison serves as a consistency check for the correct usage of the two tools. It also shows the equivalence of the \POWHEGBOX{} and the \texttt{MG5-full} implementation after the application of VBS cuts despite more approximations being used in the matrix elements entering the \POWHEGBOX{} than the \texttt{MG5-full} implementation.

For the comparison presented in this subsection we use a fixed
factorisation and renormalisation scale of
\begin{equation}
\mu_F = \mu_R = m_W/\sqrt{1+(\Gamma_W/m_W)^2}.
\end{equation} 
As before, we consider proton collisions at the LHC with
$\sqrt{s}=13$~TeV. For the selection of signal events we proceed along
similar lines as in the previous subsection.  We impose the cuts of
eqs.~(\ref{eq:jet-cuts1})-(\ref{eq:gap-cut}) defined above. 

\begin{figure}[t!]
\centering
\subfloat[][]{
\includegraphics[width=0.5\textwidth]{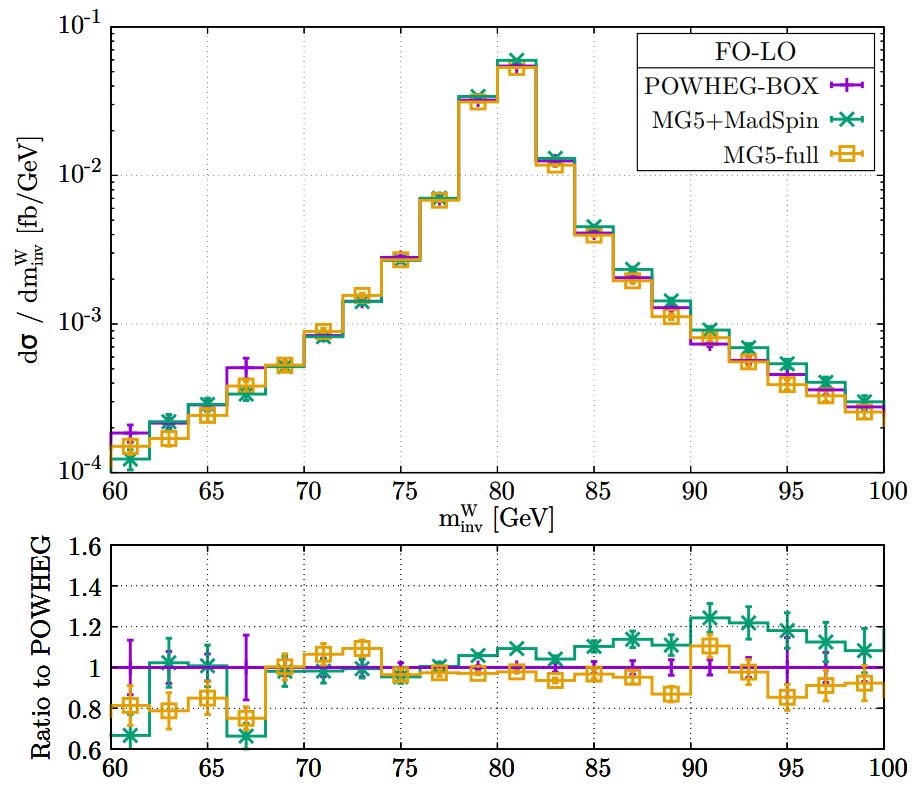}}
\subfloat[][]{
\includegraphics[width=0.5\textwidth]{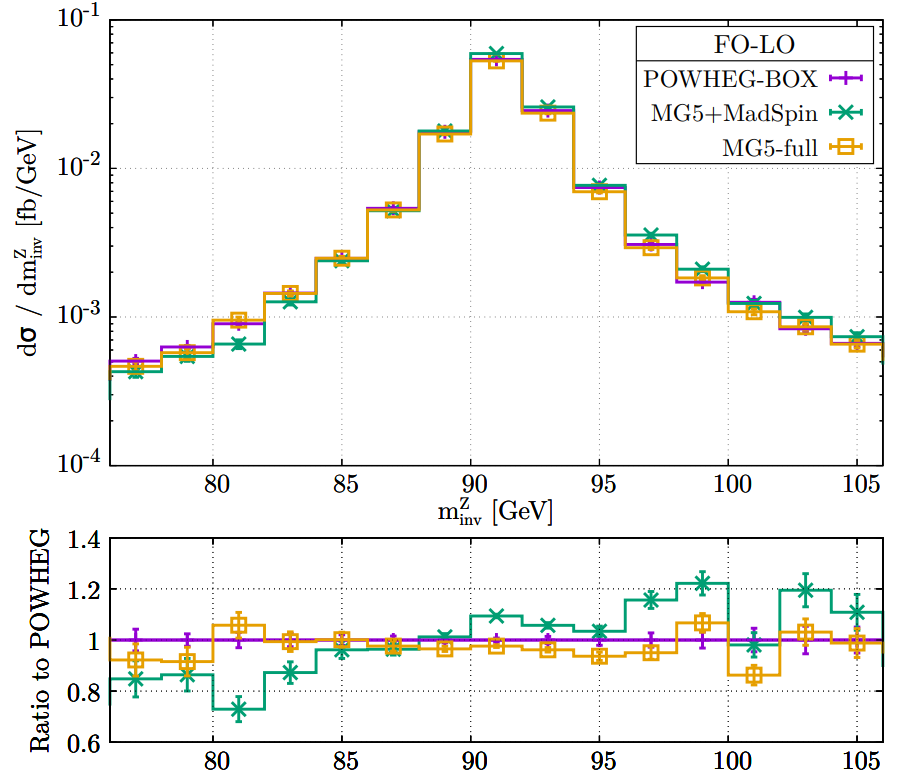}}
\qquad
\subfloat[][]{
\includegraphics[width=0.5\textwidth]{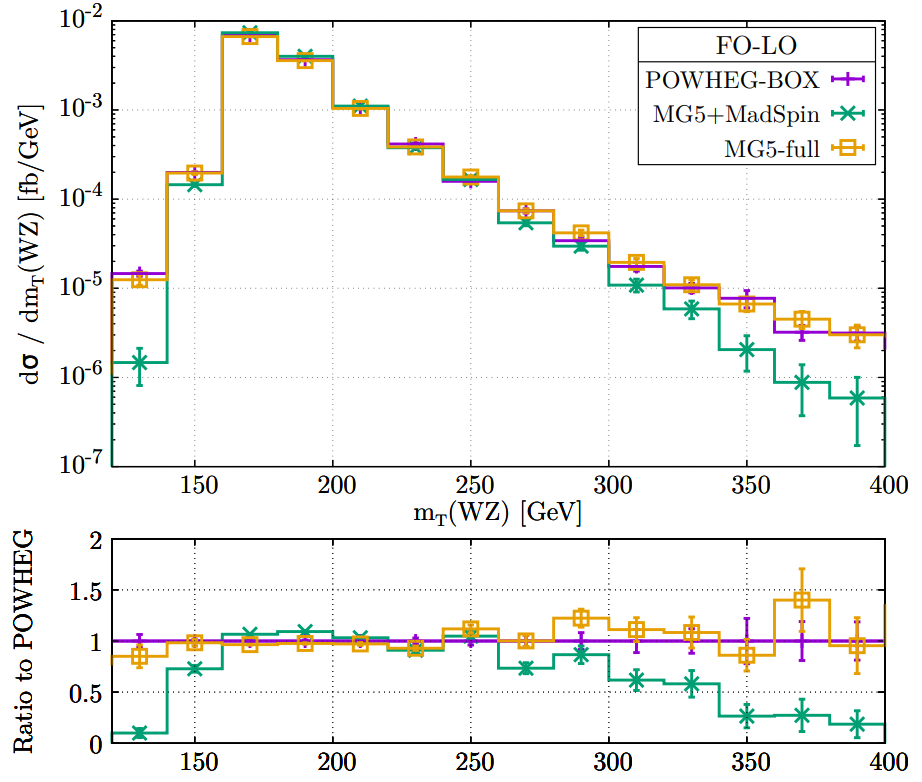}}
\subfloat[][]{
\includegraphics[width=0.5\textwidth]{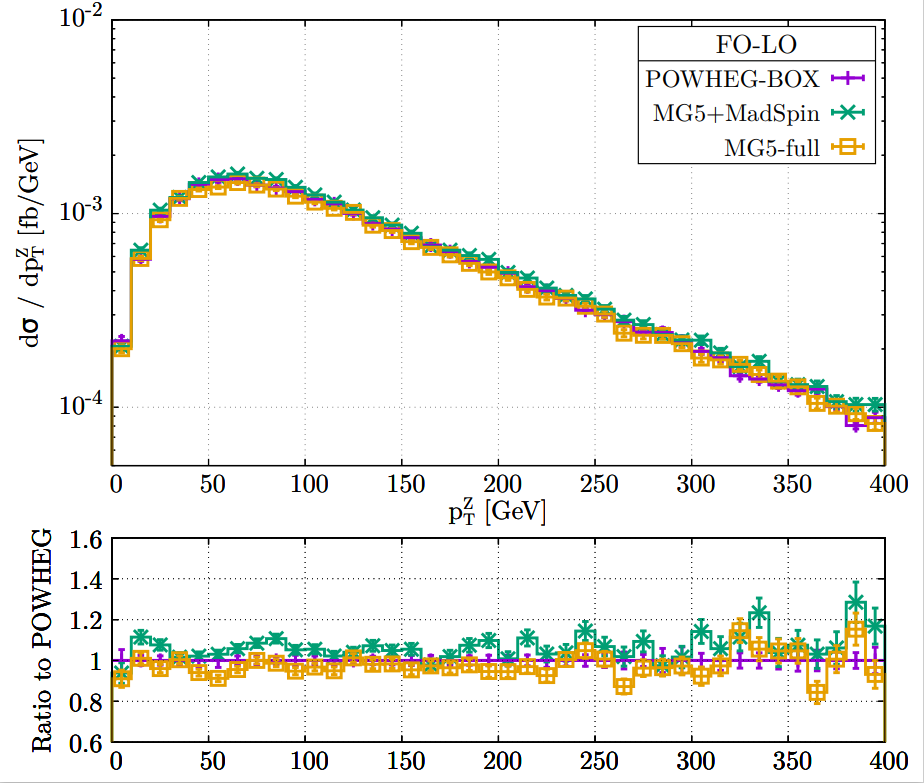}}
\caption{\label{fig:lep_LO} LO predictions for $pp\to \pmmvjj$ at the
  LHC with $\sqrt{s}=13$~TeV within the cuts of
  eqs.~(\ref{eq:jet-cuts1})-(\ref{eq:gap-cut}) as obtained with the
  \texttt{POWHEG-BOX} (purple), \texttt{MG5+MadSpin} (green), and with
  \texttt{MG5-full} (orange).
The upper panels show the reconstructed invariant mass of the $W$
boson (a), the reconstructed invariant mass of the $Z$ boson (b), the
transverse mass of the $WZ$ system (c), and the reconstructed
transverse momentum of the $Z$ boson (d). The respective lower panels
show the ratios of the \texttt{MG5+MadSpin} and \texttt{MG5-full}
predictions to the \texttt{POWHEG-BOX} results.  }
\end{figure}

In fig.~\ref{fig:lep_LO} 
%
%%%%%%%
%
%
%%%%%%
%
we show several distributions related to the decay system of the
VBS-induced $\pmmvjj$ production process. The invariant masses of the
$W$ and $Z$ systems, $m_\mr{inv}^W$ and $m_\mr{inv}^Z$, are
reconstructed from the momenta of the $\nu_e e^+$ and $\mu^-\mu^+$
pairs, respectively. The transverse mass of the $WZ$ system is defined
by
\begin{equation}
\label{eq:mtwz}
m_T(WZ) = \sqrt{(\tilde{E}_{T,W}+\tilde{E}_{T,Z})^{2} - (p_{T,W}+p_{T,Z})^{2}},
\end{equation}
where $ p_{T,W} $ and $ p_{T,Z} $ are the transverse momenta of the
reconstructed $W$ and $Z$ systems, and the transverse energies
$\tilde{E}_{T,i} $ ($i = {W,Z}$) are given by
\begin{equation}
\label{eq:et}
\tilde{E}_{T,i}=\sqrt{(m_\mr{inv}^i)^{2}+p_{T,i}^{2}}\,.
\end{equation}

We find that, for each considered distribution, the results of the
\texttt{MG5-full} and the \POWHEGBOX{} implementations are in very good agreement within their respective statistical uncertainties.
In contrast, the results of \texttt{MG5+MadSpin} deviate
from the implementations that retain full control on off-shell
contributions and spin correlations in the leptonic decay
system. While near the $W$ and $Z$ resonances \texttt{MG5+MadSpin}
yields satisfactory results, further away from the peaks of the
invariant mass distributions, and in particular in the tails of the
transverse mass distribution, the on-shell approximation does no
longer accurately reproduce the full results. Deviations can
  reach almost an order of magnitude for $m_T(WZ) \lesssim 150$~GeV
  and $m_T(WZ) \gtrsim 350$~GeV. 
This should be kept in mind for the simulation of VBS processes when
off-shell regions are of interest.

%%%%%%%%

\subsection{Semi-leptonic and hadronic decays}
While in ref.~\cite{Jager:2018cyo} an implementation for VBS-induced
$WZ$ production with fully leptonic decays was developed and made
available in the \POWHEGBOX{} program package, semi-leptonic and fully
hadronic final states were not considered before in that framework. We
have closed this gap and are now able to simulate all possible decay
modes of the $WZ$ system in the VBS mode. 

As an example for a semi-leptonic decay mode in this subsection we
provide phenomenological results for the VBS process where a $W^+$
boson decays hadronically into an $u\bar d$ pair, and the $Z$ boson
into a muon pair.  All off-shell diagrams giving rise to the same
final state are taken into account, see fig.~\ref{fig:diagrams}.

The decay quarks of the $W$ boson give rise to jets. For the selection
of the VBS signal in the presence of background processes it is
important to distinguish these decay jets from the tagging jets of the
production process. The representative numerical analysis below is
designed to take that into account.

Throughout this subsection we use the dynamical scale of
eq.~(\ref{eq:scales}).
For the selection of events in the semi-leptonic mode at the LHC with
an energy of $\sqrt{s}=13$~TeV we impose the following cuts: Charged
leptons are required to fulfill the basic requirements
\begin{equation}
\label{eq:lcuts-sh}
p_{T,\ell} > 20 \, \mathrm{GeV}, \qquad \vert y_\ell \vert < 2.47\,.
\end{equation} 
For the hardest lepton we additionally request 
\beq
 p_{T,\ell}^\mr{hardest} > 28 \, \mathrm{GeV} \,.
 \eeq
Furthermore, the invariant mass reconstructed from the muon pair has to fulfill 
\begin{equation}\label{eq:mz window slp}
83 \, \mathrm{GeV} < m_\mr{inv}^Z < 99 \, \mathrm{GeV}.
\end{equation} 
Jets and the leptons have to be separated by 
\beq 
R_{j\ell} > 0.4 \,.
\eeq 
We require all  jets to have 
\begin{equation}
 p_{T,j} > 20 \, \mathrm{GeV}, \qquad  \vert y_j \vert < 4.5\,.
\end{equation}
The jets in an event are then further classified.  The decay jets are
identified as those two jets with the invariant mass being closest to
the mass of the $W$~boson.  These two jets are required to lie within
a window around the $W$ mass of
\begin{equation}
\label{eq:decjets}
64 \, \mathrm{GeV} < m_{inv}^\mr{W} < 106 \, \mathrm{GeV}\,,
\end{equation}
and to fulfill the transverse-momentum requirements 
\begin{equation}
\label{eq:decjets-pt}
p_{T,j_1}^\mr{dec} > 40 \,  \mathrm{GeV}, \qquad 
p_{T,j_2}^\mr{dec} > 30 \, \mathrm{GeV},
\end{equation}
where $j_1$ and $j_2$ denote the hardest and second hardest decay jets.

After the selection of the decay jets, the two hardest remaining jets
are identified as the tagging jets.  For the tagging jets we require
\begin{equation}
\label{eq:tag-sh}
p_{T,j}^\mr{tag} > 30 \,  \mathrm{GeV}, \qquad 
m_{jj}^\mr{tag} > 400 \, \mathrm{GeV}, \qquad
R_{jj} > 0.4,\qquad
\eta_{j_1}^\mr{tag}\cdot \eta_{j_2}^\mr{tag} < 0\,.
\end{equation}

%
%%%%%%
%
\begin{figure}[tp!]
\centering
\subfloat[][]{
\includegraphics[width=0.5\textwidth]{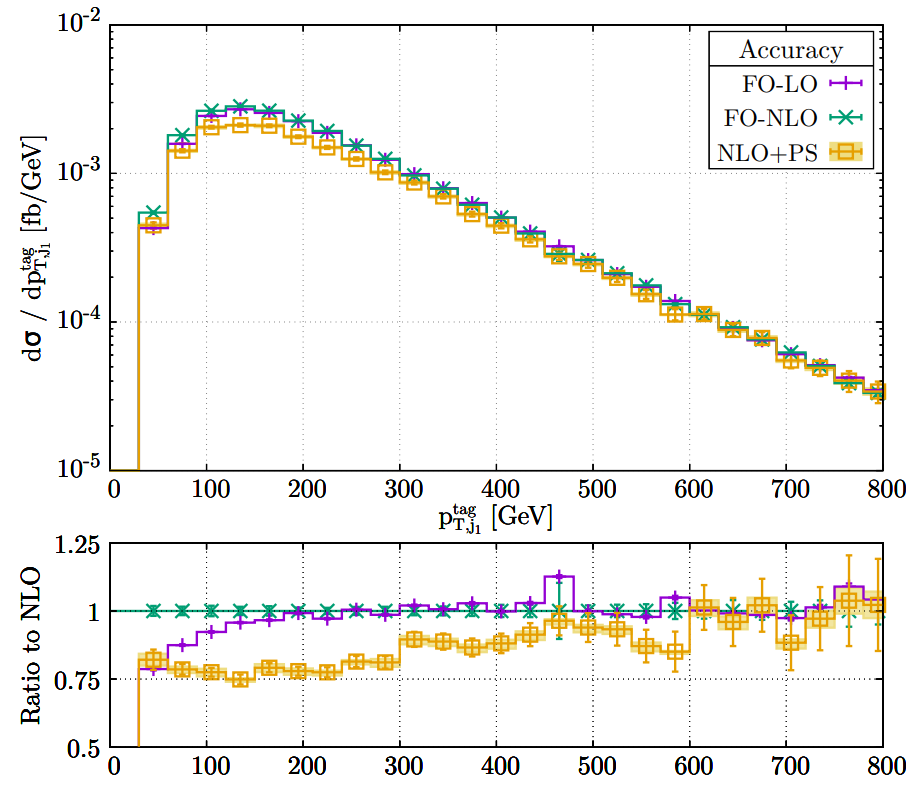}}
\subfloat[][]{
\includegraphics[width=0.5\textwidth]{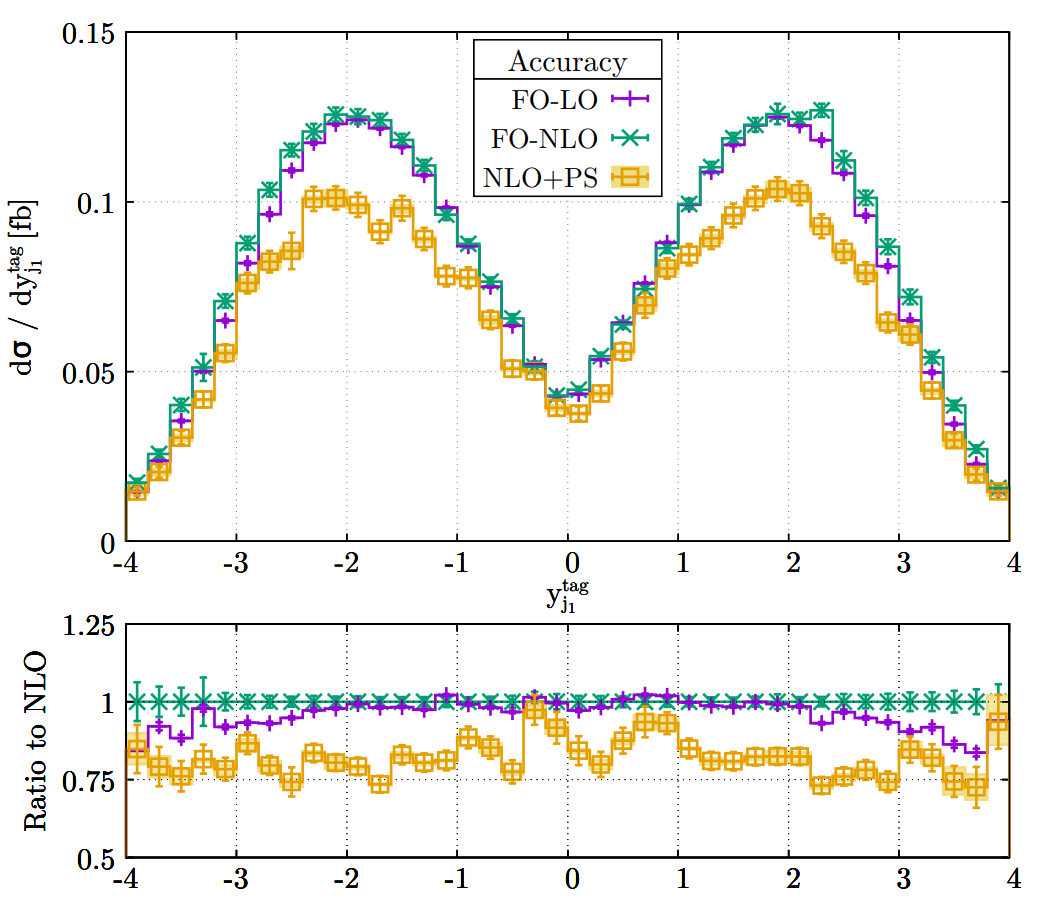}}
\qquad
\subfloat[][]{
\includegraphics[width=0.5\textwidth]{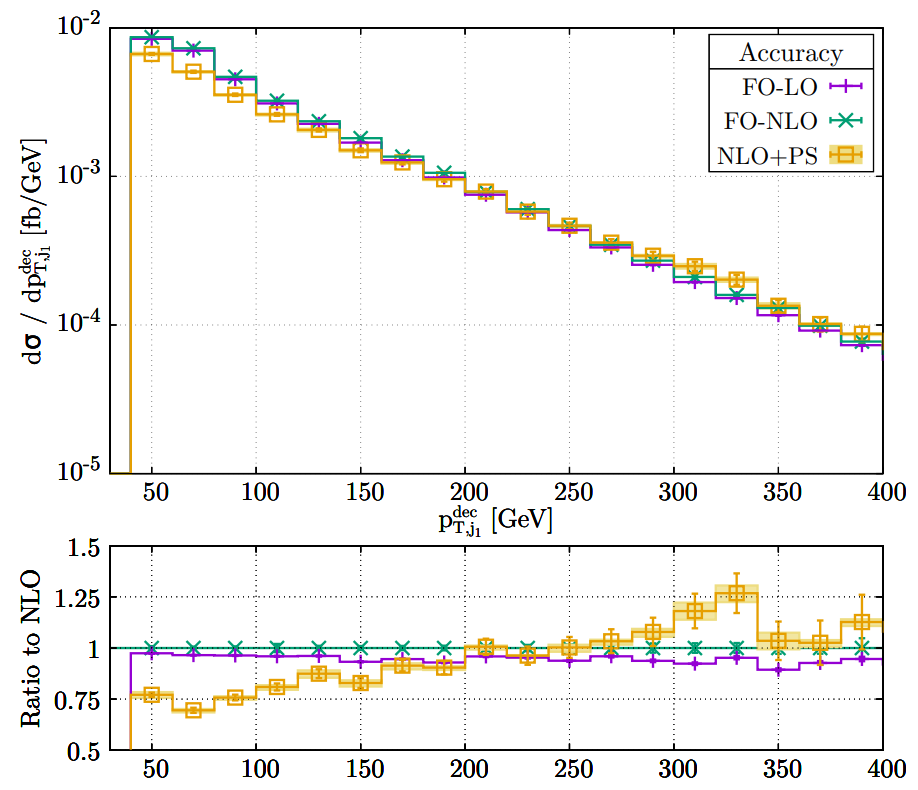}}
\subfloat[][]{
\includegraphics[width=0.5\textwidth]{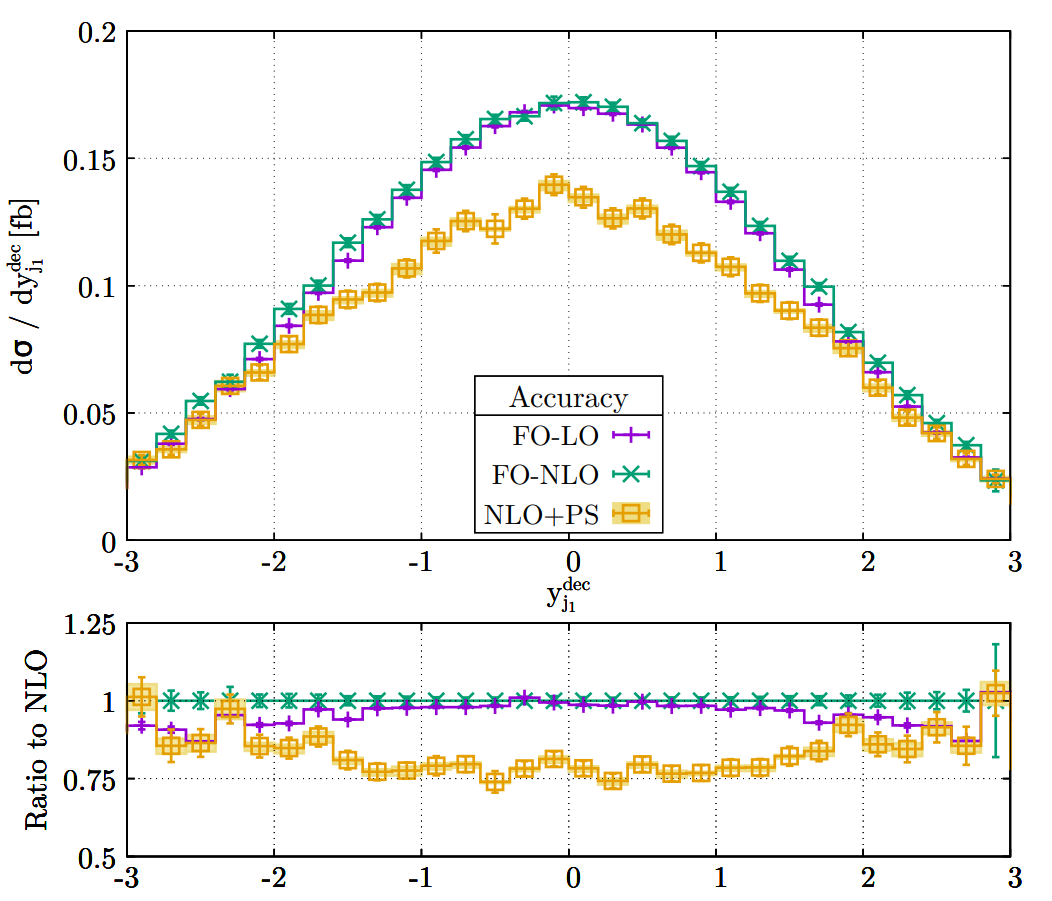}}
\caption{\label{fig:slp_fig1} Predictions for VBS-induced $W^+ Z$
  production in the semi-leptonic decay mode at the LHC with
  $\sqrt{s}=13$~TeV within the cuts of
  eqs.~(\ref{eq:lcuts-sh})--(\ref{eq:tag-sh}) at LO (purple), NLO
  (green) and NLO+PS (orange).
The upper panels show the transverse momentum of the hardest tagging
jet (a), the rapidity of the hardest tagging jet (b), the transverse
momentum of the hardest decay jet (c), the rapidity of the
hardest decay jet (d).
The respective lower panels show the ratios of the LO and NLO+PS
predictions to the NLO results.}
\end{figure}

\begin{figure}[tp!]
\centering
\subfloat[][]{
\includegraphics[width=0.5\textwidth]{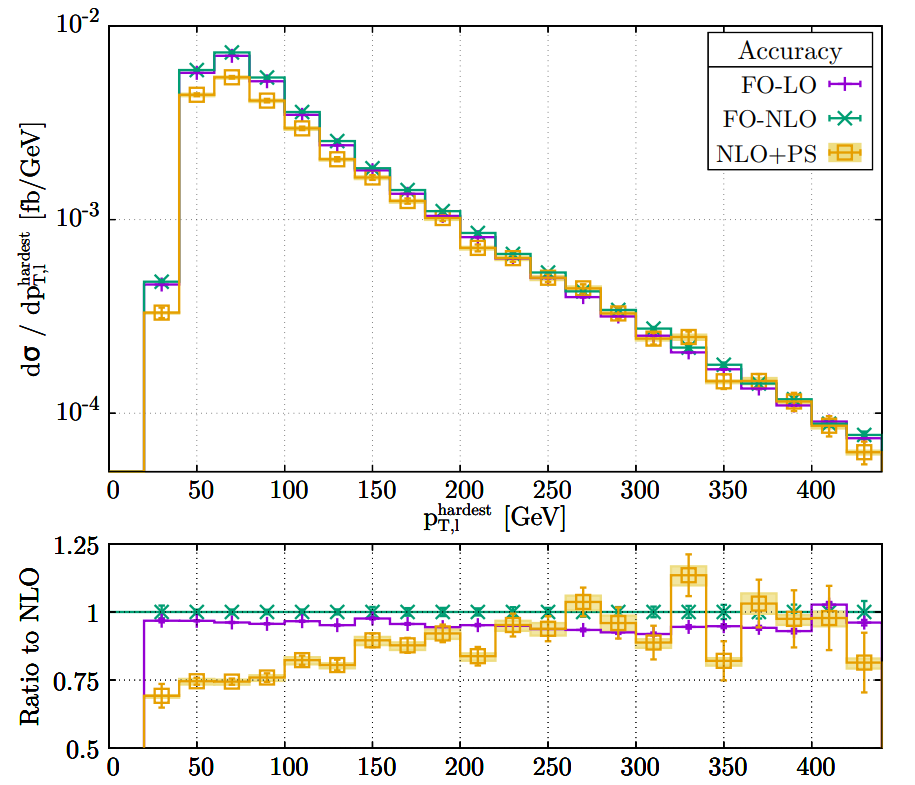}}
\subfloat[][]{
\includegraphics[width=0.5\textwidth]{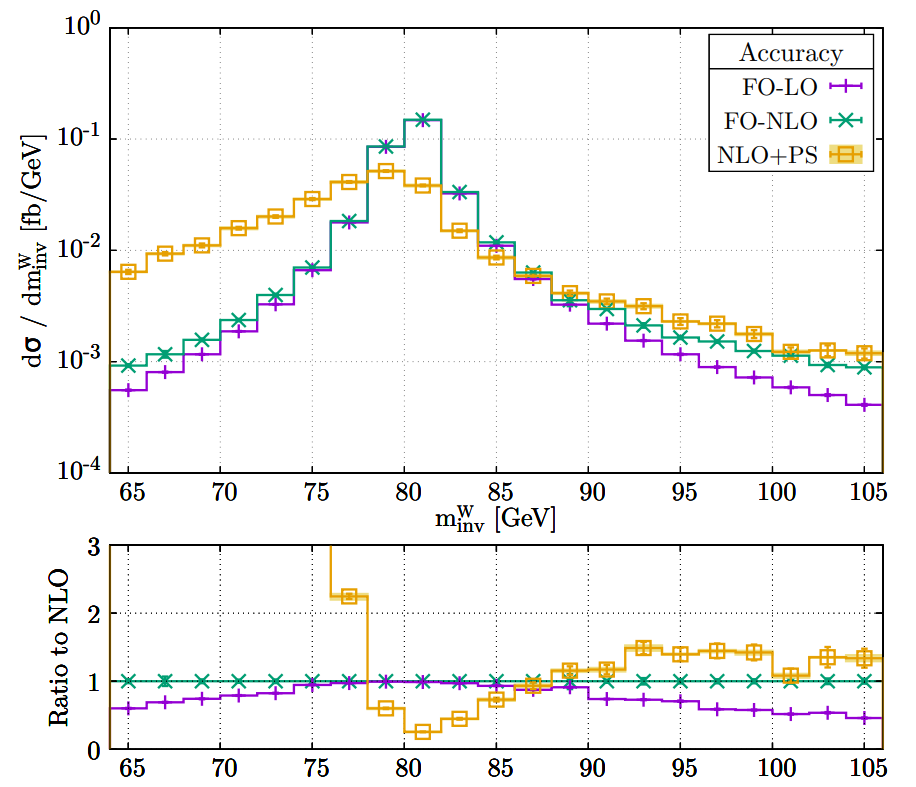}}
\caption{\label{fig:slp_fig2} Predictions for VBS-induced $W^+ Z$
  production in the semi-leptonic decay mode at the LHC with
  $\sqrt{s}=13$~TeV within the cuts of
  eqs.~(\ref{eq:lcuts-sh})--(\ref{eq:tag-sh}) at LO (purple), NLO
  (green) and NLO+PS (orange).
The upper panels show the transverse momentum of the hardest lepton
(a) and the reconstructed invariant mass of the $W$-boson system (b).
The respective lower panels show the ratios of the LO and NLO+PS
predictions to the NLO results.}
\end{figure}

%
%%%%%%
%
Fig.~\ref{fig:slp_fig1} and fig.~\ref{fig:slp_fig2} display some representative distributions
related to the tagging jets and the decay jets at LO, NLO, and NLO+PS
accuracy. As expected, tagging jets and decay jets exhibit entirely
different properties. 
Both, transverse momentum and rapidity distributions look very different for the two types of jets. From the rapidity distributions one can understand that the decay jets are preferentially located a central rapidities, while the tagging jets peak in the forward and backward regions, analogous to the leptonic decay mode.  
We note that generally
the curves for the NLO+PS results lie below the fixed-order
predictions, indicating smaller event rates. As can be deduced from
the invariant mass distribution in fig.~\ref{fig:slp_fig2}~(b) the
shift of momenta beyond LO results in a considerable change of shape
in $m_\mr{inv}^W$. However this quantity is used as a selection
criterion (c.f.\ eq.~(\ref{eq:decjets})). With $m_{inv}^\mr{W}$ shifted
to values further away from $m_W$, at NLO+PS level fewer events pass
the selection criterion of eq.~(\ref{eq:decjets}) resulting in a smaller
value of the associated cross section.

Let us now consider a representative fully hadronic decay mode with
the $W^+$ boson decaying into a $u\bar d$ pair and the $Z$ boson into
a pair of $s$ quarks. The large number of jets emerging from this mode
requires a dedicated analysis allowing for a proper classification of
tagging and decay jets.

We require all jets to have
\begin{equation}
\label{eq:jets-fh}
p_{T,j} > 10\, \mathrm{GeV}, \qquad \vert y_j \vert < 4.5\,.
\end{equation}
To identify the jets associated with the decay of the $W$ and $Z$ boson we proceed in the following way:
In the first step, from all pairs of jets we choose the one with its invariant mass closest to $m_W$. These two jets are then considered to correspond to the decay of the $W$~boson. 
In a second step, we proceed analogously for the $Z$~boson (replacing $m_W$ with $m_Z$).
In a third step, we check if a jet is contained in both the jet-pair associated with the $W$~boson and the one associated with the $Z$~boson.
If this is the case we assign it to the boson with mass closer to the corresponding jet-pair. 
In a fourth step, we repeat the first or second step for the boson that was not chosen in the third step with the remaining jets not assigned to the other boson.
The jets having thus been associated with the $W$ and $Z$ decays have
to fulfill the transverse-momentum requirements of \bea
\label{eq:decjets-fh}
p_{T,j_1}^\mr{W-dec} > 20 \,  \mathrm{GeV}, &&\qquad p_{T,j_2}^\mr{W-dec} > 10 \, \mathrm{GeV}\,,
\\\nonumber
p_{T,j_1}^\mr{Z-dec} > 20 \,  \mathrm{GeV}, &&\qquad p_{T,j_2}^\mr{Z-dec} > 10 \, \mathrm{GeV}.
\eea
and exhibit invariant masses close to the respective gauge-boson mass, 
\bea
\label{eq:minv-fh}
64 \, \mathrm{GeV} < m_{jj}^\mr{W-dec} < 88 \, \mathrm{GeV},
\\\nonumber 
84 \, \mathrm{GeV} < m_{jj}^\mr{Z-dec} < 106 \, \mathrm{GeV}.
\eea
If an event does not pass these cuts, it is discarded.

After the selection of the decay jets, the two hardest remaining jets
are identified as the tagging jets.  As in the semi-leptonic case, for
the tagging jets we additionally require

\begin{equation}
\label{eq:tags-fh}
p_{T,j}^\mr{tag} > 30 \,  \mathrm{GeV}, \qquad 
m_{jj}^\mr{tag} > 400 \, \mathrm{GeV}, \qquad
R_{jj} > 0.4, \qquad 
\eta_{j_1}^\mr{tag}\cdot \eta_{j_2}^\mr{tag} < 0\,.
\end{equation}

%%%%%%%%%%%%

\begin{figure}[pt!]
\centering
\subfloat[][]{
\includegraphics[width=0.5\textwidth]{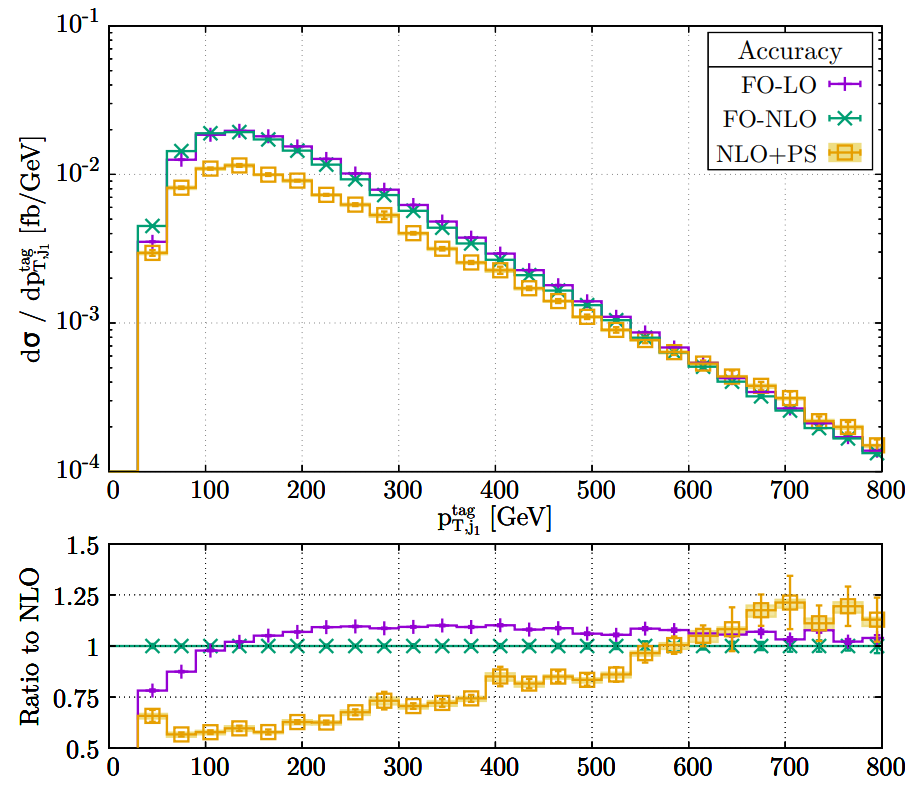}}
\subfloat[][]{
\includegraphics[width=0.5\textwidth]{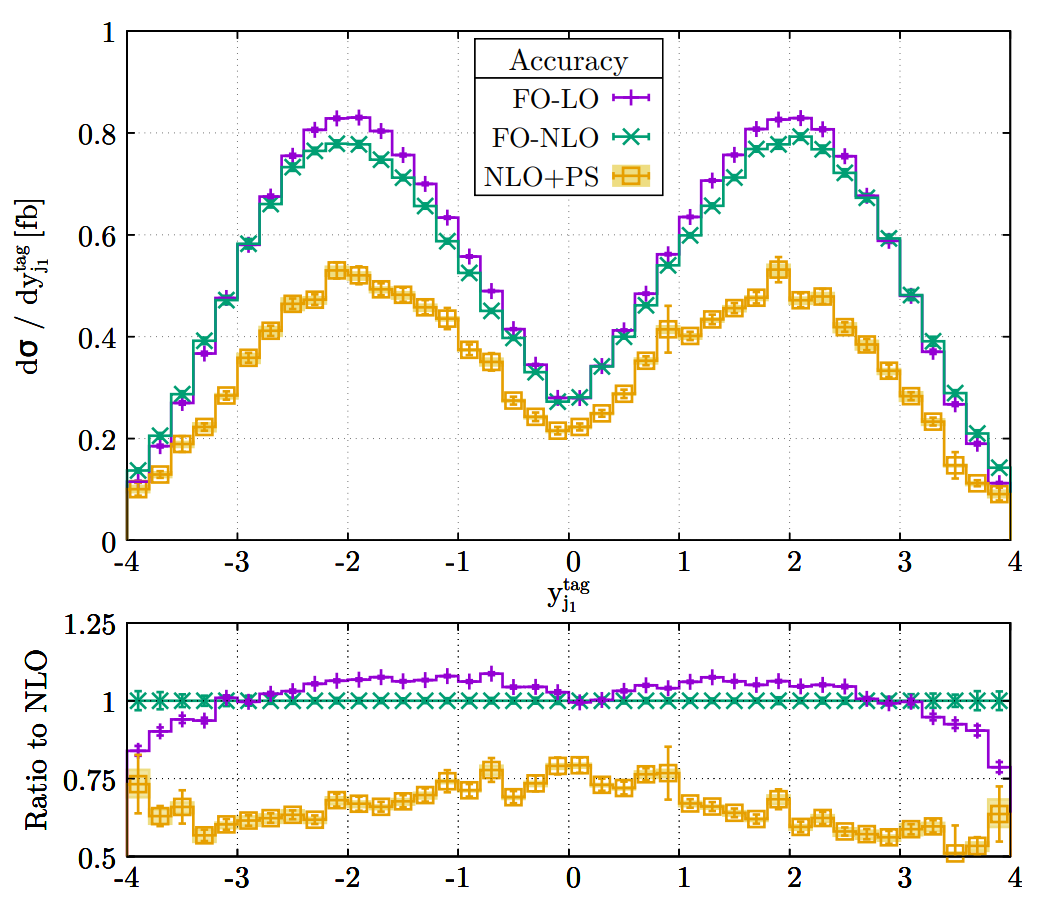}}
\qquad
\subfloat[][]{
\includegraphics[width=0.5\textwidth]{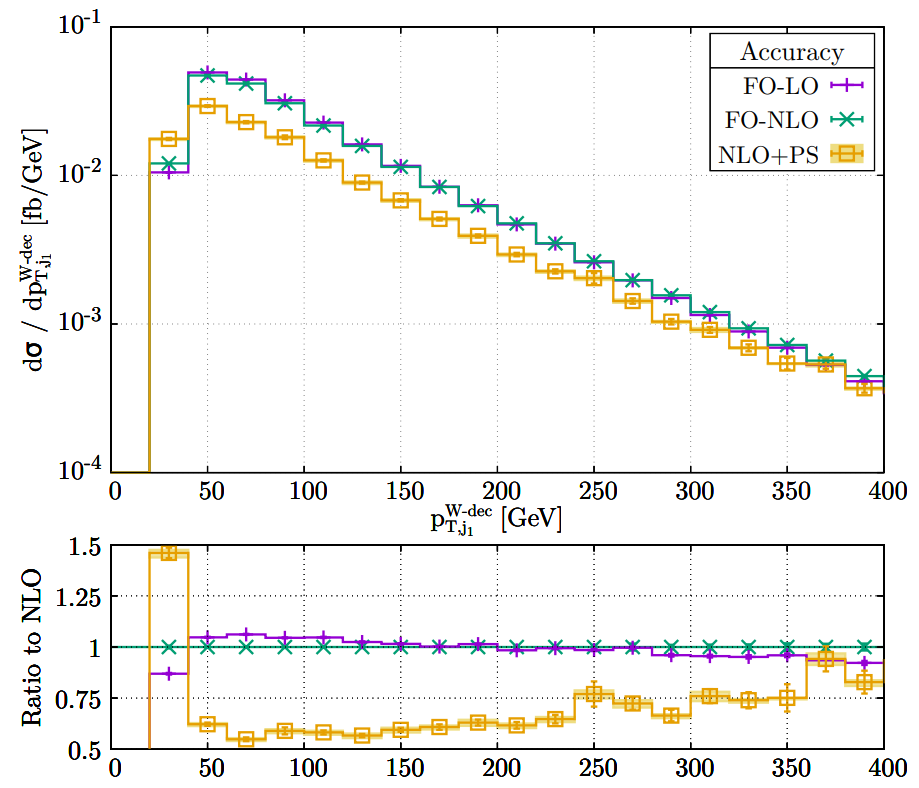}}
\subfloat[][]{
\includegraphics[width=0.5\textwidth]{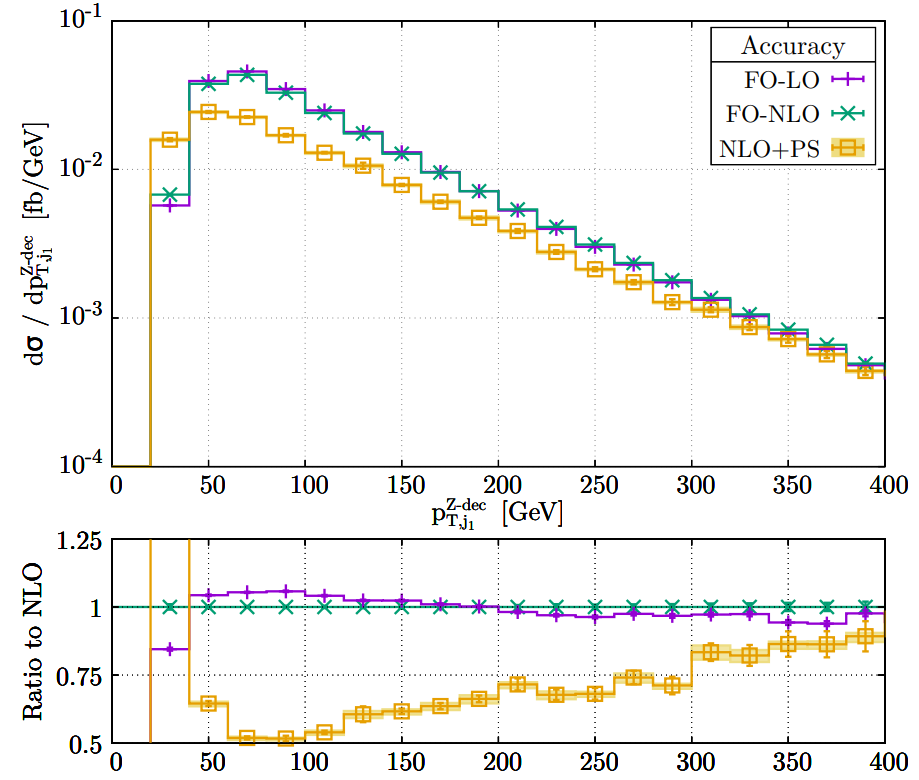}}
\caption{\label{fig:had1} Predictions for VBS-induced $W^+ Z$
  production in the fully hadronic decay mode at the LHC with
  $\sqrt{s}=13$~TeV within the cuts of
  eqs.~(\ref{eq:jets-fh})--(\ref{eq:tags-fh}) at LO (purple), NLO
  (green) and NLO+PS (orange).
The upper panels show the transverse momentum of the hardest tagging
jet (a), the rapidity of the hardest tagging jet (b), the transverse
momentum of the hardest decay jet from the $W$ decay (c), the
transverse momentum of the hardest decay jet from the $Z$ decay (d).
The respective lower panels show the ratios of the LO and NLO+PS
predictions to the NLO results.  }
\end{figure}

\begin{figure}[pt!]
\centering
\subfloat[][]{
\includegraphics[width=0.5\textwidth]{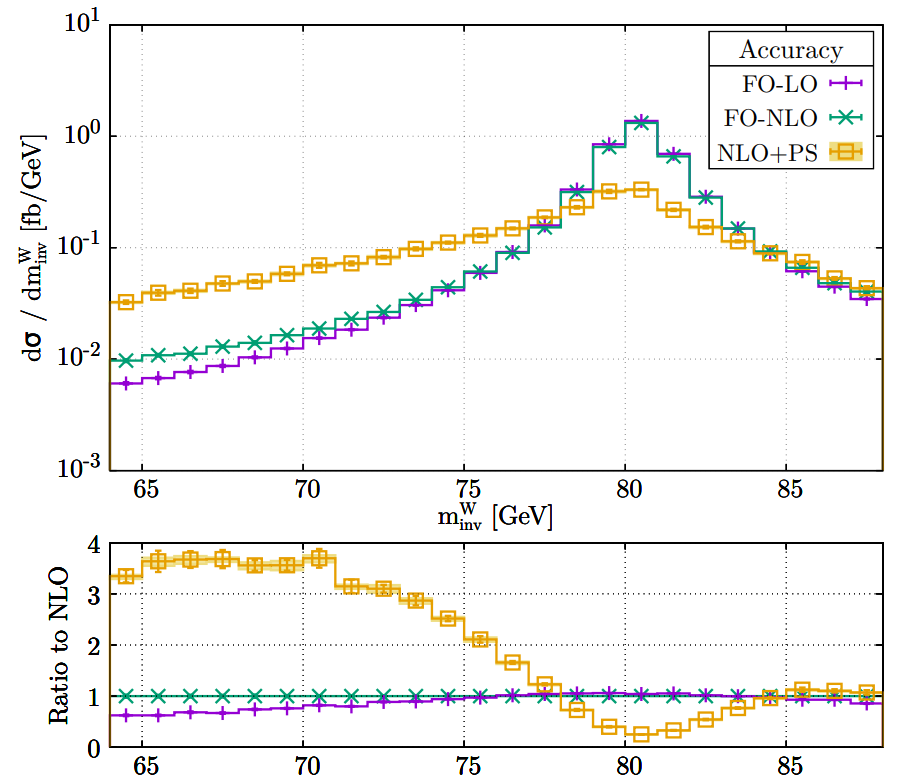}}
\subfloat[][]{
\includegraphics[width=0.5\textwidth]{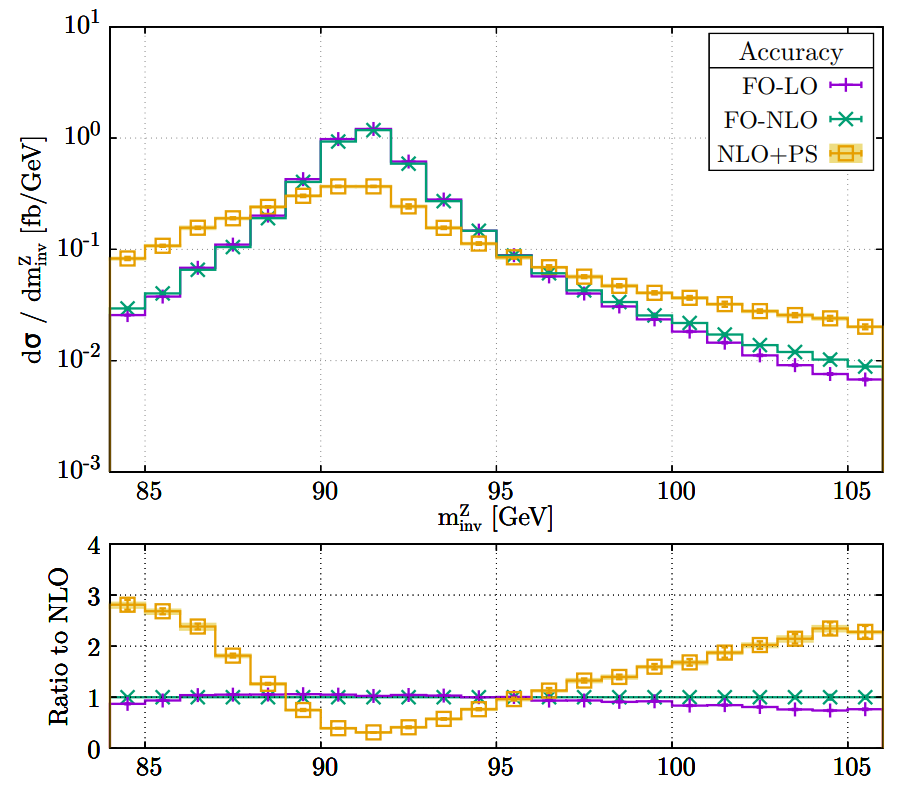}}
\caption{\label{fig:had2} Predictions for VBS-induced $W^+ Z$
  production in the fully hadronic decay mode at the LHC with
  $\sqrt{s}=13$~TeV within the cuts of
  eqs.~(\ref{eq:jets-fh})--(\ref{eq:tags-fh}) at LO (purple), NLO
  (green) and NLO+PS (orange).
The upper panels show the reconstructed mass of the $W$ boson (a), the reconstructed mass of
the $Z$ boson (b).
The respective lower panels show the ratios of the LO and NLO+PS
predictions to the NLO results.  }
\end{figure}

In fig.~\ref{fig:had1} and fig.~\ref{fig:had2} we show several distributions for the fully
hadronic decay mode.

%
%%%%%%%%%%
%
We observe results that are qualitatively similar to the semi-leptonic
case discussed above. However, the ``smearing effect'' in the NLO+PS
results is even larger than in the semi-leptonic case, because now
window cuts for both the $Z$ and the $W$ system have to be passed for
an event to be accepted.

\section{Conclusions and outlook}
\label{sec:conclusions}
In this article we have presented new features for the implementation
of VBS-induced $WZjj$ production in the framework of the
\POWHEGBOXVV. We are providing semi-leptonic and fully hadronic decays
of the intermediate vector bosons that were missing in the previously
existing implementation, and account for physics beyond the SM by the
inclusion of dimension-six operators of a generic EFT expansion in the
EW sector.

To illustrate the capabilities of the updated implementation we
considered some selected applications. We explored the sensitivity of
typical VBS observables to dimension-six operators in an EFT
framework, and found that, when the expansion is performed
consistently, predictions with contributions from EFT operators
compatible with current experimental limits barely deviate from the SM
case at LHC energies. Larger effects are found at higher energies
which could be achieved, for instance, at a future FCC.

Using the leptonic decay mode as an example, we investigated the
relevance of including off-shell contributions and spin correlations
in the simulation. By comparing our results to those obtained with the
\MGAMCNLO +\texttt{MadSpin} tool that combines a calculation of
VBS-induced $WZ$ production with a simulation of the gauge-boson
decays we could show that the on-shell approximation is appropriate
when all final-state leptons stem from the resonant decay of a gauge
boson, but deviates from the full result in regions away from the
resonance. This limitation of the approximation should be kept in mind
for ensuring its application is restricted to its region of validity.

Finally, we considered semi-leptonic and fully hadronic decay
modes. We found that in these cases QCD corrections and PS effects can
lead to a reshuffling of momenta such that they pass different
selection cuts than the corresponding LO configurations. This
kinematic effect results in a reduction of cross section beyond the
LO, which becomes particularly pronounced once the NLO result is
matched with a PS.

The new features of the VBS $\wzjj$ code have been made available via the \POWHEGBOXVV{} repository, see \url{https://powhegbox.mib.infn.it/}.

%=================================================
%
\section*{Acknowledgements}
%
%=================================================
We are grateful for valuable discussions with Johannes Scheller.  We appreciate helpful input from our experimental colleagues, in particular Joany Manjarres. This work has been supported by the German Federal Ministry for Education and Research (BMBF) under contract no.~05H21VTCAA. 
The authors acknowledge support by the state of Baden-W\"urttemberg
through bwHPC and the German Research Foundation (DFG) through grant
no INST 39/963-1 FUGG. 

%=================================================
%

\bibliographystyle{JHEP}
\bibliography{wzjj}

\end{document}